\documentclass[final]{siamltex}

\usepackage{epsfig,amssymb,latexsym}
\usepackage{amsfonts,psfrag,amsmath,bbm,color}
\usepackage{cancel}
\usepackage{siunitx}
\usepackage{comment}
\usepackage{mathrsfs}
\usepackage{graphicx}
\usepackage{textcomp}
\usepackage{multirow}
\usepackage{enumerate}
\usepackage{cancel}
\usepackage{algpseudocode}
\usepackage{caption}  
\usepackage{subcaption}
\usepackage{url}
\usepackage{rotating}
\usepackage{bigints}
\usepackage{hyperref}
\usepackage{xcolor}
\usepackage{ulem}
\usepackage{mathrsfs}
\usepackage{setspace}

\usepackage{placeins}

\usepackage{algorithm}
\usepackage{geometry}
\spacing{1.075}
\vbadness=\maxdimen

\newcommand{\footremember}[2]{%
    \footnote{#2}
    \newcounter{#1}
    \setcounter{#1}{\value{footnote}}%
}
\newcommand{\footrecall}[1]{%
    \footnotemark[\value{#1}]%
} 
\usepackage{comment}
\usepackage{tikz}
\usetikzlibrary{shapes.geometric, arrows}
\tikzstyle{startstop} = [rectangle, rounded corners, minimum width=1cm, minimum height=1cm,text centered, draw=black]
\tikzstyle{io} = [trapezium, trapezium left angle=70, trapezium right angle=110, minimum width=1cm, minimum height=1cm, text centered, draw=black, fill=blue!30]
\tikzstyle{method} = [rectangle, rounded corners, minimum width=1cm, minimum height =1cm, text centered, draw=black]
\tikzstyle{process} = [rectangle, minimum width=1cm, minimum height=1cm, text centered, draw=black]
\tikzstyle{decision} = [diamond, minimum width=0.5cm, minimum height=0.5cm, text centered, draw=black, fill=green!30]
\tikzstyle{arrow} = [thick,->,>=stealth]

\usepackage{amscd}

\graphicspath{{./figs/}}

\newtheorem{remark}{Remark}[section]

\def\PP{{{\rm l}\kern - .15em {\rm P} }}
\def\PN2{{\PP_{N}-\PP_{N-2}}}

\newcommand{\A}{\mathbbm{A}}
\newcommand{\B}{\mathbbm{B}}

\newcommand{\E}{\mathbbm{E}}

\newcommand{\Y}{\mathbbm{Y}}

\newcommand{\cD}{\mathcal{D}}

\newcommand{\bphi}{\boldsymbol{\varphi}}

\newcommand{\bpsi}{\boldsymbol{\psi}}
\newcommand{\btau}{\boldsymbol{\tau}}

\newcommand{\bif}{\textbf{\textit{f}}\hspace{-0.5mm}}
\newcommand{\ba}{\boldsymbol{a}}

\newcommand{\bH}{\boldsymbol{H}}

\newcommand{\bL}{\boldsymbol{L}}

\newcommand{\bu}{\boldsymbol{u}}

\newcommand{\bv}{\boldsymbol{v}}

\newcommand{\bw}{\boldsymbol{w}}
\newcommand{\bW}{\boldsymbol{W}}

\newcommand{\btu}{\tilde{\boldsymbol{u}}}

\newcommand{\bx}{\boldsymbol{x}}
\newcommand{\bX}{\boldsymbol{X}}

\newcommand{\bY}{\boldsymbol{Y}}
\newcommand{\by}{\boldsymbol{y}}

\newcommand{\bGamma}{\boldsymbol{\Gamma}}

\newcommand{\red}[1]{{\color{red}#1}}

\newcommand{\deleted}[1]{{}}

\hypersetup{hidelinks}

\begin{document}
\title{Data-driven correction Reduced-Order Stochastic Collocation Method for the Time-Dependent Navier-Stokes Equations}

\author{
Md. Abdul Aziz\footremember{uabm}{D\MakeLowercase{epartment of} M\MakeLowercase{athematics}, U\MakeLowercase{niversity of} A\MakeLowercase{labama at} B\MakeLowercase{irmingham}, AL 35294, USA; P\MakeLowercase{artially supported by the} N\MakeLowercase{ational} S\MakeLowercase{cience} F\MakeLowercase{oundation} (NSF) \MakeLowercase{grant} DMS-2425308.}%
\and Muhammad Mohebujjaman\footnote{C\MakeLowercase{orrespondence: mmohebuj@uab.edu}}\hspace{1mm}\footrecall{uabm}
\and  Xuping Xie\footremember{odu}{D\MakeLowercase{epartment of} M\MakeLowercase{athematics and } S\MakeLowercase{tatistics}, O\MakeLowercase{ld} D\MakeLowercase{ominion} U\MakeLowercase{niversity}, N\MakeLowercase{orfolk}, VA 23529, USA.}
 }
\maketitle

\begin{abstract}
    We develop two hybrid data-driven Uncertainty Quantification (UQ) Reduced-Order Models (ROMs) for the non-stationary incompressible Navier--Stokes Equations (NSEs) with random viscosity. The proposed framework combines Proper Orthogonal Decomposition (POD) ROMs, spatial filtering, data-driven closure modeling, and the Stochastic Collocation Method (SCM). A key feature is the approximation of the realization-dependent closure term by the closure associated with the expected viscosity. Consequently, a single data-driven closure model constructed from the full-order solution corresponding to the expected viscosity can be reused across all realizations, thereby avoiding the expensive construction of realization-specific closure models. This leads to the Data-Driven Correction UQ ROM (DDC-UQ-ROM) and its physically constrained variant, the Constrained DDC-UQ-ROM (CDDC-UQ-ROM), in which physical constraints are imposed on the data-driven correction model.

The proposed methods are evaluated for a two-dimensional (2D) flow past a circular cylinder problem with a 5D random viscosity field, considering mean Reynolds numbers $Re=100$ and $Re=500$, different correlation lengths, and both in-sample and out-of-sample parameter settings. Numerical results show that the DDC-UQ-ROM generally improves upon the standard Galerkin ROM (G-ROM), while the CDDC-UQ-ROM exhibits the most accurate and robust approximation of the benchmark energy evolution across the cases considered. The robustness and accuracy of the CDDC-UQ-ROM are further assessed through statistical error analyses of the first- and second-order energy moments, reduced-training-data experiments, and an indicative comparison of wall-clock computational costs with the full-order UQ calculations.
\end{abstract}



\pagestyle{myheadings}
\thispagestyle{plain}

\markboth{\MakeUppercase{Data-Driven Correction Stochastic Collocation ROM for Navier–Stokes Equations}}{\MakeUppercase{  M. A. Aziz, M. Mohebujjaman, and X. Xie}}

\section{Introduction} Many fluid-flow models depend on physical parameters that are not known exactly because of measurement errors, material variability, unresolved physical processes, or other sources of uncertainty. Here, we consider the random viscosity and study the following stochastic time-dependent incompressible Navier--Stokes equations (NSEs) \cite{gunzburger2019evolve}:\setlength{\abovedisplayskip}{0pt} \setlength{\abovedisplayshortskip}{0pt}
\setlength{\belowdisplayskip}{0pt} \setlength{\belowdisplayshortskip}{0pt}
\begin{equation}
\left\{\begin{aligned}
    \bu_t + \bu \cdot \nabla \bu - \nabla\cdot\left(\nu(\bx, \omega) \nabla \bu\right) + \nabla p &= \bif(\bx, t), &\forall (\bx, t) \in \cD \times (0, T], \\
    \nabla \cdot \bu &= 0, &\forall (\bx, t) \in \cD \times (0, T], \\
    \bu &= \textbf{g}(\bx), &\forall (\bx, t) \in \partial \cD \times (0, T], \\
    \bu(\bx, 0) &= \bu_0(\bx), &\forall \bx \in \cD,
\end{aligned}\right.\label{strong-form-problem}
\end{equation}where $\cD \subset \mathbb{R}^s$, $s=2,3$, is a bounded domain
with sufficiently smooth boundary $\partial\cD$, and $T>0$ denotes the simulation end time. The temporal and spatial variables are represented by $t$ and $\bx$, respectively. Consider a complete probability space $(\Omega,\mathscr{F},P)$.
Here, $\Omega$ represents the collection of possible outcomes,
$\mathscr{F}\subseteq 2^\Omega$ is a $\sigma$-algebra, and
$P:\mathscr{F}\to[0,1]$ assigns probabilities to measurable events. The flow velocity and pressure are denoted by $\bu$ and $p$, respectively, and $\bif$ is the body force. For simplicity, we consider $\textbf{g}(\bx)=\textbf{0}$. The random viscosity is represented by $\nu(\bx,\omega)$ with the dependence on $\omega\in\Omega$. The resulting solution $(\bu(\bx,t,\omega),p(\bx,t,\omega))$ is a random field on $\cD\times[0,T]\times\Omega$.  This modeling choice is natural in settings where viscosity is uncertain due to temperature‑dependent rheology, mixture composition, impurities, calibration and measurement errors, or unresolved subgrid dissipation. Because viscosity directly controls dissipation, randomness in $\nu$ perturbs the effective Reynolds number $Re$ and the production–dissipation balance, thereby altering transition thresholds, intermittency, and the long‑time statistics of coherent structures. The principal computational challenge is that the UQ for \eqref{strong-form-problem} requires propagating {distributions} over $\nu$ through a high‑fidelity flow solver. Intrusive polynomial‑chaos or stochastic Galerkin discretizations couple all modes and can lead to very large algebraic systems \cite{Powell2009,Sousedik2016,Sousedik2022};

To compute flow ensemble, for each fixed realization $\omega\in\Omega$, we obtain the following deterministic variable‑coefficient NSEs: \setlength{\abovedisplayskip}{0pt} \setlength{\abovedisplayshortskip}{0pt}
\setlength{\belowdisplayskip}{0pt} \setlength{\belowdisplayshortskip}{0pt}
\begin{equation}
\left\{\begin{aligned}
\bu_{j,t}+\bu_j\cdot\nabla \bu_j-\nabla\cdot\left(\nu_j(\bx) \nabla \bu_j\right)+\nabla p_j &=  \bif,\\
	\nabla\cdot \bu_j & = 0,
\end{aligned}\right.\label{gov2-rans}
\end{equation}where $\bu_j$, $p_j$, and $\nu_j$ denote, respectively, the velocity, pressure, and viscosity corresponding to the $j^{th}$ realization. A standard approach in UQ \cite{berry2025efficient, islam2026efficient, jiang2024second} approximates \eqref{strong-form-problem} by solving \eqref{gov2-rans} for $j=1,2,\ldots,J$, and taking the ensemble average, where $J$ is the total number of realizations and $J\gg 1$. 

Monte Carlo methods (MCMs) and their variants generally require numerous independent realizations to obtain reliable statistical estimates, in contrast, SCMs \cite{babuska2007stochastic,nobile2008sparse,xiu2005high} requires far fewer realizations. Although nonintrusive surrogate modeling and SCMs avoid modifications to existing deterministic solvers, constructing accurate approximations of stochastic flow responses can still require numerous high-fidelity full-order simulations. Each of  these full-order simulations is itself expensive for moderate–high Reynolds numbers because fine spatial resolution and small time steps are needed to capture the nonlinear energy cascade and boundary layers. Consequently, the total cost can be prohibitive when long time ensemble simulations are involved.

Combining ROMs with SCMs may address this computational burden. ROMs project the high-dimensional system dynamics onto a lower-dimensional space constructed from coherent structures extracted from data \cite{antoulas2005approximation}, often using POD \cite{holmes2012turbulence,noack2011reduced}. However, standard Galerkin ROMs (G-ROMs) may exhibit reduced accuracy and stability in convection-dominated regimes due to the neglected interactions between resolved and unresolved flow scales. To address these limitations, several stabilization and data-driven closure approaches have been developed \cite{ivagnes2023hybrid,Ivagnes2023Pressure,mohebujjaman2019physically,mohebujjaman2017energy,xie2018data} for deterministic NSEs. These approaches have demonstrated the potential of learned correction terms to improve the predictive performance of ROMs.

For stochastic NSEs, Resseguier et al. \cite{resseguier2021quantifying} developed a stochastic POD-ROM framework for quantifying uncertainties associated with truncated flow scales, while Mou et al. \cite{mou2023efficient} proposed a data-driven multiscale stochastic ROM for complex nonlinear systems. These stochastic closure approaches address modeling uncertainty in reduced dynamics and are therefore distinct from approaches that propagate uncertainty in prescribed physical parameters. A stabilized evolve-filter-relax reduced-order SCM was developed in \cite{gunzburger2019evolve} to provide stable and efficient UQ approximations. Their approach demonstrated the effectiveness of higher-order differential filters in ROMs of convection-dominated stochastic flows. 

Despite these advances, developing accurate and computationally efficient data-driven ROMs for UQ of parameter-dependent NSEs remains challenging. In particular, incorporating learned closure corrections into stochastic collocation frameworks requires careful treatment of parameter variability, unresolved-scale effects, and physical constraints. These considerations motivate the investigation of DDC-UQ-ROM and their physically constrained counterparts CDDC-UQ-ROM for random-parameter NSEs flows, with emphasis on predictive accuracy, computational efficiency, and the approximation of statistical quantities. To the best of our knowledge, the DDC-UQ-ROM and CDDC-UQ-ROM are novel for UQ-ROM of fluid flow problems.

The rest of the paper is organized as follows. In Section \ref{ddc-uq-roms-snse}, we describe the proposed data-driven correction ROMs. The weak formulation and fully discrete full-order model of the stochastic NSEs are presented in Section \ref{weak-snse}. A brief discussion of the SCMs is provided in Section \ref{scm-uq}. The POD G-ROMs and their basis construction are described in Section \ref{G-ROM-construction}. In Section \ref{numerical-exp}, we present numerical results for the proposed data-driven ROM closure models applied to 2D flow past a circular cylinder with 5D random viscosity, yielding random Reynolds numbers with mean values of $Re=100$ and $Re=500$. Finally, conclusions and directions for future research are presented in Section \ref{conclusion}.

\section{Data-Driven Correction UQ-ROMs for Fluid Flows}
\label{ddc-uq-roms-snse}
In this section, we first describe the standard G-ROMs for \eqref{gov2-rans}, followed by the proposed data-driven correction ROMs, namely, the DDC-UQ-ROM and CDDC-UQ-ROM.

\subsection{G-ROMs or Projection ROMs (Proj-ROMs)} \label{proj-ROM} To build G-ROMs, in projection‑based approaches such as POD-Galerkin \cite{ballarin2015supremizer}, one constructs the basis from snapshots of high‑fidelity solutions and advances the flow via a small system of Ordinary Differential Equations (ODEs) for the modal coefficients. Hyper‑reduction, e.g., Empirical Interpolation Method (EIM), the Discrete Empirical Interpolation Method (DEIM) or gappy POD~\cite{peherstorfer2020stability,willcox2006unsteady}, is often employed to reduce the online cost of evaluating nonlinear and parameter‑dependent operators. Projection ROMs (Proj-ROMs) are among the most popular classes of ROMs. From the modes $\{\bphi_1,\ldots,\bphi_d\}$, Proj-ROMs select the dominant modes $\{\bphi_1,\ldots,\bphi_r\}$, with $r\le d$, as the ROM basis functions that capture the recurrent spatial structures, where $d$ denotes the rank of the snapshot matrix. It then replaces $\bu$ in the governing Partial Differential Equations (PDEs) with the Galerkin truncation $\bu_r = \sum\limits_{i=1}^r a_i \bphi_i$, and subsequently projects the resulting PDE system onto the ROM space $\bX_r := \text{span}\{\bphi_1, \dots, \bphi_r\}$. As a result, the parameterized Proj-ROM for \eqref{gov2-rans} yields the following low-dimensional dynamical system:  \begin{align}
        \dot{\ba_j} = \A \ba_j + \ba_j^\top \B \ba_j,\;\;j=1,2,\cdots,J, \label{eqn-1.1}
    \end{align}
    where $\ba_j\in \mathbb{R}^r$ denotes the vector of ROM coefficients, and $\A \in \mathbb{R}^{r \times r}$, $\B \in \mathbb{R}^{r \times r \times r}$ are ROM operators. The ROM operators are computed in the offline stage.

In practical applications, Proj-ROM \eqref{eqn-1.1} often produces inaccurate results \cite{amsallem2012stabilization,holmes2012turbulence} when a small number of modes is used. This is because the Proj-ROM is not closed and thus fails to capture interactions between resolved and unresolved modes \cite{aubry1988dynamics,couplet2003intermodal}. In addition to deterministic closure and stabilization approaches, stochastic ROMs have also been used to represent uncertainty associated with truncated unresolved scales \cite{resseguier2021quantifying}. Thus, for realistic flow simulations, the following modified Proj-ROM is used:
\begin{align}
    \dot{\ba_j} = \A \ba_j + \ba_j^\top \B \ba_j + \btau_j,\;\;j=1,2,\cdots,J, \label{eqn-1.2}
\end{align}
where $\btau_j$ models the interaction between resolved modes $\{\bphi_1, \dots, \bphi_r\}$ and unresolved modes $\{\bphi_{r+1}, \dots, \bphi_d\}$. A classical approach is to approximate $\btau_j$ using an Eddy Viscosity (EV) as below \cite{wang2012proper}: 
\begin{equation}
    \btau_j \approx \text{EV}(\ba_j). \label{eqn-1.3}
\end{equation}

\subsection{Data-driven ROMs (DD-ROMs)} The family of purely DD-ROMs \cite{brunton2016discovering,kutz2016dynamic,peherstorfer2015dynamic,peherstorfer2016data,schmid2010dynamic} is also written in the same form of \eqref{eqn-1.1} but computes the operators $\A$ and $\B$ in a least-squares sense minimizing the residual between the DD-ROM trajectories and Full Order Model (FOM) or experimental data. More recently, data-driven operator inference has been extended to
parameterized PDEs, enabling nonintrusive ROMs for
many-query applications involving varying physical parameters
\cite{mcquarrie2023nonintrusive}. DD-ROMs often solve nonlinear inverse problems, such as finding low-dimensional operators, which are generally sensitive to small perturbations in the training data, leading to high numerical instability \cite{alla2026efficient,beinhorn2025solving,borcea2024data}.

\subsubsection{Data-driven filtered ROM} Although classical ROMs can achieve orders‑of‑magnitude speed ups, naive projections of \eqref{strong-form-problem} with random viscosity frequently misrepresent how truncated scales modulate dissipation. In particular, (i) the resolved energy may drift due to an under‑resolved energy transfer from neglected modes; (ii) the mean and variance of key observables can be biased across viscosity realizations; and (iii) stationary statistics may be distorted when the reduced viscous operator fails to reflect the spatial variability in $\nu(\bx,\omega)$. These issues are exacerbated when the snapshot basis is built at a nominal viscosity and then reused far from that nominal value. To overcome these limitations, a hybrid projection/data-driven ROM (Proj-DD-ROM) \cite{ivagnes2023hybrid,mohebujjaman2019physically,xie2018data} is introduced to exploit the complementary strengths of the two approaches. In particular, Ivagnes et al. developed hybrid data-driven closure strategies for deterministic NSEs that combine physics-based and data-driven closure modeling for reduced-order fluid-flow models \cite{ivagnes2023hybrid}. In this framework, the operators $\A$ and $\B$ in \eqref{eqn-1.2} are obtained through projection, whereas the unresolved contribution $\btau_j$ is identified from data. The resulting ROM, which is then called the DDC-UQ-ROM, is conceptually represented as follows:
\begin{align}
\text{FOM filtering + projection} \longrightarrow \text{F-ROM} \longrightarrow \text{DDC-UQ-ROM}, \label{eqn-1.4}
\end{align}
where the arrows indicate the transition steps and F-ROM stands for filtered ROM. More recently, data-driven closure strategies have been extended to parametrized ROMs, where the closure is learned as a function of both the reduced state and the underlying physical parameters \cite{ivagnes2026data}. In contrast, in the present stochastic setting, we approximate the realization-dependent closure by the closure associated with the expected viscosity so that a single closure model can be reused across all stochastic realizations.

\subsection*{Step 1: Projection Approach}
The first stage of the DDC-UQ-ROM \eqref{eqn-1.4} constructs the reduced operators $\A$ and $\B$ in \eqref{eqn-1.2} together with an explicit representation of the closure term $\btau_j$. For this purpose, a ROM spatial filter is applied to the FOM solution to extract its resolved large-scale components. The influence of the unresolved scales on the resolved dynamics is represented through the closure term $\btau_j$. This procedure yields a F-ROM, which describes the large-scale flow dynamics using fewer modes than the FOM. The resulting F-ROM is written as
\begin{align}
    \dot{\ba_j} = \mathbb{A}\ba_j + \ba_j^\top \mathbb{B} \ba_j + \btau_j ,\;\;j=1,2,\cdots,J, \label{eqn-1.5}
\end{align}
where $\btau_j$ has an explicit dependence on the FOM data.

We note that the F-ROM \eqref{eqn-1.5} cannot be used directly until the closure problem is addressed, namely, until an appropriate expression for $\btau_j$, $j=1,2,\ldots,J$, is specified. This term is computed in the offline stage but is computationally expensive and often infeasible to compute for all realizations. Thus, we approximate $\btau_j$ by $\btau_{\Bar{\nu}}$, where $\btau_{\Bar{\nu}}$ is the ROM closure term corresponding to $\mathbb{E}[\nu]$ and is independent of the realizations. In this paper, we derive an exact mathematical formula for $\btau_{\Bar{\nu}}$ using a ROM projection spatial filter. This filter seeks the orthogonal projection of $\bu \in \bX_h$ onto the ROM space $\bX_r$ in $L^2$-sense, for a fixed $r \le d$, by finding $\overline{\bu}_r \in \bX_r$ such that
\begin{align}
(\bu - \overline{\bu}_r, \bphi_i) = 0, \quad \forall i = 1,2,\cdots,r,
\end{align}
where $\bX_h$ is a finite element space. The projection filtered ROM for NSE is obtained as below \cite{mohebujjaman2019physically,xie2018data}:
\begin{align}
\left(\btu_{r,t},\bphi_i\right)+\left(\btu_r\cdot\nabla\btu_r,\bphi_i\right)+\left(\mathbb{E}[\nu]\nabla\btu_r,\nabla\bphi_i\right)+\left(\btau_{\Bar{\nu}},\bphi_i\right)=0,\quad \forall i = 1,2,\cdots,r,\label{filtered-nse}
\end{align}
where the subfilter-scale stress tensor \begin{align}
\btau_{\Bar{\nu}}=\overline{\btu_d\cdot\nabla\btu_d}^r-\btu_r\cdot\nabla\btu_r,\label{sub-filter-scale-stress-tensor}
\end{align} $\btu$ is the solution of the NSE corresponding to the viscosity $\mathbb{E}[\nu]$, $\overline{\btu}$ denotes the spatial filtering of $\btu$, and $\btu_d$ denotes the spatial filtered approximation of $\btu$ in the $\bX_d$.

Note that, in \eqref{filtered-nse}, we assume that the spatial filter and the differential operator commute \cite{sagaut2006large}. A more detailed discussion of the commutation error in ROMs can be found in \cite{koc2019commutation}.


The validity of the approximation $\btau_j\approx \btau_{\Bar{\nu}}$ is assessed primarily through its effect on the resulting ROM solutions rather than by explicitly computing \(\btau_j\) for every realization. In particular, although the closure model is constructed using the FOM solution corresponding to \(\mathbb{E}[\nu]\), the resulting data-driven ROMs are evaluated over a substantially larger set of stochastic collocation points \cite{nobile2008sparse,stoyanov2015tasmanian,stoyanov2016dynamically}, including out-of-sample parameter realizations. The accurate reproduction of the benchmark energy evolution and its statistical moments over these realizations provides a posteriori evidence that the expected-viscosity closure captures the dominant unresolved-scale effects over the parameter range considered. That is, we discard the effect of the small order term $\btau_j-\btau_{\Bar{\nu}}$ in the system.

\subsection*{Step 2: DDC-UQ-ROM}
To obtain a closed F-ROM while maintaining a form similar to the Proj-ROM in  \eqref{eqn-1.1}, we follow \cite{xie2018data} and introduce the following quadratic ansatz:
\begin{align}
    \btau_{\Bar{\nu}}^{ansatz} = \widetilde{\A}\ba_j + \ba_j^\top \widetilde{\B} \ba_j, \label{eqn-1.7-new}
\end{align}
where the operators $\widetilde{\A}\in \mathbb{R}^{r \times r}$, and $\widetilde{\B}\in \mathbb{R}^{r \times r\times r}$ are obtained via data-driven modeling. To find $\widetilde{\A}$ and $\widetilde{\B}$, the low-dimensional unconstrained minimization problem given below is solved, which minimizes the difference between $\btau_{\Bar{\nu}}$ computed from the FOM data and $\btau_{\Bar{\nu}}^{ansatz}$:
\begin{align}
    \min_{\widetilde{\A},\; \widetilde{\B}} \|\btau_{\Bar{\nu}}(\text{FOM}) - (\widetilde{\A}\ba_j + \ba_j^\top \widetilde{\B}\ba_j)\|^2. \label{eqn-1.8}
\end{align}
Here, $\|\cdot\|$ denotes the Euclidean norm. To solve the unconstrained minimization problem, FOM snapshot data corresponding to $\E[\nu]$ are used to construct an overdetermined linear system, which is then solved in the least-square sense. That is, we want to find $\widetilde{\A}$, and $\widetilde{\B}$ as \begin{align}
\min_{\substack{\widetilde{\A}\in\mathbb{R}^{r\times r}\\\widetilde{\B}\in\mathbb{R}^{r\times r\times r}}} 
\sum_{n=1}^{M}
\left\|
\btau_{\Bar{\nu}}(\text{FOM})(t_n)
-
\left(
\widetilde{\A}\ba_{\Bar{\nu}}^{snap}(t_n) + \ba_{\Bar{\nu}}^{snap}(t_n)^{T}\widetilde{\B}\ba_{\Bar{\nu}}^{snap}(t_n)
\right)
\right\|^2,\label{time-uncos}
\end{align}
where the coordinate vector $\ba_{\Bar{\nu}}^{snap}(t_n)$ is computed by projecting the snapshot at $t=t_n$ corresponding to $\E[\nu]$ onto the POD basis. $M$ is the total number of snapshots used to solve the above least-square problem. By collecting all unknown entries of $\widetilde{\A}$ and $\widetilde{\B}$ into a single vector $\bx\in\mathbb{R}^{r^2+r^3}\hspace{-1mm},$ and assembling the corresponding data matrix in $F\in\mathbb{R}^{(Mr)\times(r^2+r^3)}$ and subfilter-scale stress tensor data into vector $\bif\in\mathbb{R}^{Mr}$, the optimization problem \eqref{time-uncos} can be written as 
\begin{equation}
\min_{\bx} \| \bif - F \bx \|^2.
\label{eq:linear_LS}
\end{equation}
The least-squares problem may be solved through the Singular Value Decomposition (SVD) of $F$, written as $F=U\Sigma V^T$. Here, $U$ and $V$ are orthogonal matrices, while $\Sigma$ is rectangular and diagonal, with the nonnegative singular values $\sigma_i$ appearing on its diagonal. However, the overdetermined system $F\bx=\bif$ is highly ill-conditioned because some of its singular values are close to zero. To obtain a stable solution, we construct a low-rank approximation of $F$ using a Truncated SVD (\textit{TSVD}) algorithm with a prescribed tolerance $tol_d$, namely,
\[
\hat{U},\hat{\Sigma},\hat{V}=TSV\hspace{-0.6mm}D(F,tol_d),
\]
as described in Algorithm~\ref{truncated-svd}. Then, the least-squares solution is approximated by
\[
\bx=\hat{V}\hat{\Sigma}^{-1}\hat{U}^T\bif.
\]
Finally, the resulting DDC-UQ-ROM is given by the following dynamical system:
\begin{align}
    \dot{\ba}_j = (\A + \widetilde{\A})\ba_j + \ba_j^\top (\B + \widetilde{\B})\ba_j.
    \label{eqn-1.9}
\end{align}
\begin{algorithm}
	\caption{Truncated SVD: $\hat{U},\hat{\Sigma},\hat{V}=TSV\hspace{-0.6mm}D(E,tol)$}\label{truncated-svd}
	\begin{algorithmic}[1]
		\State \textbf{Initialization:} Given a matrix $E\in\mathbb{R}^{m\times n}$, a tolerance $tol\in\mathbb{R}^+$
		\State Determine the SVD of $E$: $E=U\Sigma V^T$
		\State Construct $\hat{\Sigma}$ from $\Sigma$ as: $\hat{\sigma}_i=\sigma_i$ if $\sigma_i>tol$
        \State Delete corresponding  columns of $U$, and $V$ to form $\hat{U}$, and $\hat{V}$, respectively
        \State\Return {$\hat{U}$, $\hat{\Sigma}$, and $\hat{V}$}
	\end{algorithmic}
\end{algorithm}

\subsubsection{Physically Constrained DDC-UQ-ROM (CDDC-UQ-ROM)}
Another class of DDC-UQ-ROM is the CDDC-UQ-ROM \cite{mohebujjaman2019physically}, in which a constrained optimization problem is solved, instead of \eqref{eqn-1.8}, to determine the operators $\widehat{\A}\in \mathbb{R}^{r \times r}$ and $\widehat{\B}\in \mathbb{R}^{r \times r\times r}$, where the quadratic ansatz: \begin{align}
    \btau_{\Bar{\nu},c}^{ansatz} = \widehat{\A}\ba_j + \ba_j^\top \widehat{\B} \ba_j. \label{anstz-const}
\end{align} The key observation is that the Proj-ROM  operator $\A$ is negative semi-definite and thus $\ba^T_j\A\ba_j \le 0$, whereas the operator $\B$ satisfies $\ba_j^T [\ba_j^T\B\ba_j ]=0$ if one uses the skew-symmetric formulation. Thus, we impose analogous constraints on $\widehat{\A}$ and $\widehat{\B}$ so that the following holds $\ba_j^T\widehat{\A}\ba_j\le 0$ and $\ba_j^T [\ba_j^T\widehat{\B}\ba_j ]=0$. That is, we replace \eqref{eqn-1.8} by
\begin{equation}
\left\{\begin{aligned}
& \underset{\widehat{\A},\; \widehat{\B}}{\min}
& & \|\btau_{\Bar{\nu}}(\text{FOM}) - (\widehat{\A}\ba_j + \ba_j^\top \widehat{\B}\ba_j)\|^2  \\
& \text{subject to}
& & \ba_j^T\widehat{\A}\ba_j\le 0,\\
&&& \ba_j^T [\ba_j^T\widehat{\B}\ba_j ]=0.
\end{aligned}\right.\label{const-optimi}
\end{equation} Using the sufficient conditions for $\ba_j^T\widehat{\A}\ba_j\le 0$ and $\ba_j^T [\ba_j^T\widehat{\B}\ba_j ]=0$, and introducing the FOM snapshot data corresponding to $\E[\nu]$, \eqref{const-optimi} can be written as follows: 
\begin{equation}
\left\{\begin{aligned}
& \min_{\substack{\widehat{A}\in\mathbb{R}^{r\times r}\\\widehat{B}\in\mathbb{R}^{r\times r\times r}}} 
&&\sum_{n=1}^{M}
\left\|
\btau_{\Bar{\nu}}(\text{FOM})(t_n)
-
\left(
\widehat{\A}\ba_{\Bar{\nu}}^{snap}(t_n) + \ba_{\Bar{\nu}}^{snap}(t_n)^{T}\widehat{\B}\ba_{\Bar{\nu}}^{snap}(t_n)
\right)
\right\|^2  \\
& \text{subject to}
& & \widehat{\A}_{ii}\le 0,\;\forall i=1,2,\cdots, r,\\
&&& \widehat{\A}_{ik}=-\widehat{\A}_{ki},\;\forall i,k=1,2,\cdots, r,\;\; i\not=k,\\
&&&\widehat{\B}_{i i i}
	=
	0, 
	\ 
	\forall \, i = 1, \ldots, r,\\
&&& \widehat{\B}_{i i k}
	+ 
	\widehat{\B}_{i k i}
	+
	\widehat{\B}_{k i i}
	=
	0, 
	\ 
	\forall \, i,k = 1, \ldots, r, \ i \neq k \, ,\\
&&& \widehat{\B}_{i k l}
		+ 
		\widehat{\B}_{i l k}
		+
		\widehat{\B}_{k i l}
		+
		\widehat{\B}_{k l i}
		+
		\widehat{\B}_{l i k}
		+
		\widehat{\B}_{l k i}
		=
		0 , 
		\  
		\forall \, i,k,l = 1, \ldots, r, \ i \neq k \neq l \neq i \, .
\end{aligned}\right.\label{const-optimi-data}
\end{equation}
Similar to the DDC-UQ-ROM, collect all unknown entries of $\widehat{\A}$ and $\widehat{\B}$ into a column vector $\widehat{\bx}$ and assemble a coefficient matrix $G$. Since $G$ is an ill-conditioned matrix, for a chosen tolerance $tol_c$ we find a low-rank approximation $\widehat{G}$ of $G$ as follows: $$\hat{U},\hat{\Sigma},\hat{V}=TSV\hspace{-0.6mm}D(G,tol_c)$$ then construct $\widehat{G}=\hat{U}\hat{\Sigma}\hat{V}^T$.

Now, introduce a small positive constant $\epsilon$ following the previous work in \cite{mohebujjaman2019physically} (which provides additional computational control),  we construct the constrained optimization problem as follows
\begin{equation}
\left\{\begin{aligned}
& \underset{\widehat{\bx}\in\mathbb{R}^{r^2+r^3}}{\min}
& & \|\bif - \widehat{G}\widehat{\bx}\|^2  \\
& \text{subject to}
& & \widehat{\A}_{ii}\le -\epsilon,\;\forall i=1,2,\cdots, r,\\
&&& \widehat{\A}_{ik}=-\widehat{\A}_{ki},\;\forall i,k=1,2,\cdots, r,\;\; i\not=k,\\
&&&\widehat{\B}_{i i i}
	=
	0 , 
	\ 
	\forall \, i = 1, \ldots, r,\\
&&& \widehat{\B}_{i i k}
	+ 
	\widehat{\B}_{i k i}
	+
	\widehat{\B}_{k i i}
	=
	0 , 
	\ 
	\forall \, i,k = 1, \ldots, r, \ i \neq k \, ,\\
&&& \widehat{\B}_{i k l}
		+ 
		\widehat{\B}_{i l k}
		+
		\widehat{\B}_{k i l}
		+
		\widehat{\B}_{k l i}
		+
		\widehat{\B}_{l i k}
		+
		\widehat{\B}_{l k i}
		=
		0 , 
		\  
		\forall \, i,k,l = 1, \ldots, r, \ i \neq k \neq l \neq i \, .
\end{aligned}\right.\label{const-optimi-final}
\end{equation} We solve \eqref{const-optimi-final} using MATLAB function \textit{lsqlin} with \textit{interior-point} algorithm, $ConstraintTolerance=10^{-9}$, $OptimalityTolerance=10^{-9}$, and $StepTolerance=10^{-9}$. Once we compute, $\widehat{\A}$, and $\widehat{\B}$, we have the following dynamical system for CDDC-UQ-ROM: \begin{align}
    \dot{\ba_j} = (\A + \widehat{\A}) \ba_j + \ba_j^\top (\B + \widehat{\B}) \ba_j. \label{eqn-1.9-cDDC-UQ-ROM}
\end{align}

\section{Weak Formulation of Stochastic NSEs and Its Fully Discrete FOM}\label{weak-snse} We next derive the variational form associated with \eqref{strong-form-problem}. Let $(\cdot,\cdot)$ denote the inner product in $L^2(\cD)$. We introduce the velocity space $\bX:=\bH_0^1(\cD)$, the pressure space $Q:=L_0^2(\cD)$, and the stochastic space $\bW:=\bL_P^2(\Omega)$. The corresponding stochastic weak formulation is: Find $\bu \in \bX \otimes \bW$ and $p \in Q \otimes \bW$ such that, for a.e. $t\in(0,T]$
\begin{equation}
\left\{\begin{aligned}
    \mathbb{E}[(\bu_t, \bv)] + \mathbb{E}[(\bu \cdot \nabla \bu, \bv)] + \mathbb{E}[ (\nu\nabla \bu, \nabla \bv)] - \mathbb{E}[(p, \nabla \cdot \bv)] &= \mathbb{E}[(\bif, \bv)], &\forall \bv \in \bX \otimes \bW, \\
    \mathbb{E}[(\nabla \cdot \bu, q)] &= 0, &\forall q \in Q \otimes \bW, \\
    \bu(\bx, 0, \omega) &= \bu_0(\bx).
\end{aligned}\right.
\end{equation}

For many practical problems, the underlying uncertainty may be approximated by a finite-dimensional set of random variables \cite{babuska2007stochastic,nobile2008sparse}. To enable efficient numerical approximation, we replace the infinite-dimensional randomness $\omega \in \Omega$ by a finite-dimensional parametric representation $\by \in \bGamma$. Introduce the random parameter vector $\by=(y_1,\ldots,y_d)\in\bGamma\subset\mathbb{R}^d$, whose joint probability density is denoted by $\rho(\by)$, with $\bGamma=\prod_{\ell=1}^d\Gamma_\ell$. Using this finite-dimensional parametrization, the viscosity field $\nu(\bx,\omega)$ is represented by $\nu(\bx,\by)$. With respect to the weight $\rho(\by)$, we further define $\bY:=\bL_\rho^2(\bGamma)$ as the weighted space of square-integrable functions over $\bGamma$. The corresponding parametric weak formulation is to find $\bu\in\bX\otimes\bY$ and $p\in Q\otimes\bY$ such that, for almost every $t\in(0,T]$, the following relations hold:
\begin{equation}
\left\{\begin{aligned}
    \int_{\bGamma}(\bu_t, \bv)\rho(\by)d\by + \int_{\bGamma} (\bu \cdot \nabla \bu, \bv)\rho(\by)d\by &+ \int_{\bGamma} (\nu(\bx, \by)\nabla \bu, \nabla \bv)\rho(\by)d\by \\
    - \int_{\bGamma} (p, \nabla \cdot \bv)\rho(\by)d\by &= \int_{\bGamma} (\bif, \bv)\rho(\by)d\by, &\forall \bv \in \bX \otimes \bY, \\
    \int_{\bGamma} (\nabla \cdot \bu, q)\rho(\by)d\by &= 0, &\forall q \in Q \otimes \bY.\label{int-weak-form}
\end{aligned}\right.
\end{equation}

To enable efficient offline/online decomposition in ROMs, we assume an affine dependence of the viscosity on the parameters:
\begin{equation}\label{eq:affine-nu}
\nu(\bx,\by)=\nu_0(\bx)+\sum_{\ell=1}^d \nu_\ell(\bx)\, y_\ell.
\end{equation}
 If the dependence of $\nu$ is nonaffine, empirical interpolation techniques, such as EIM/DEIM, Matrix Discrete Empirical Interpolation Method (MDEIM), or gappy variants ~\cite{quarteroni2016reduced} can be employed to construct an accurate affine surrogate.

 \subsection{Fully Discrete FOM}
 Let $\bX_h \subset \bX$ and $Q_h \subset Q$ be conforming Finite Element (FE) spaces for the velocity and pressure, respectively, satisfying the appropriate discrete inf-sup stability condition. To generate FOM solutions used for snapshot construction, we employ an FE discretization in space together with the second-order Backward Differentiation Formula-2 (BDF-2) for the temporal discretization.

 Let the temporal grid be uniform with step size $\Delta t$, and set $t_n=n\Delta t$, $n=0,1,\ldots,N$, where $N=T/\Delta t$. Denote $\bu_{j,h}^n$ as the FE solution at time $t_n$ for $j$-th realization. Assume the divergence-free initial conditions $\bu_{j,h}^0,\bu_{j,h}^1$ are in $\bH^2(\cD)$. The fully discrete linearized BDF-2 time-stepping scheme for solving \eqref{gov2-rans} is therefore: Find $(\bu_{j,h}^{n+1}, p_{j,h}^{n+1}) \in \bX_h \times Q_h,\;\;j=1,2,\cdots,J$ such that for all $(\bv_h, q_h) \in \bX_h \times Q_h$,
\begin{align}
\frac{1}{2\Delta t}\left(
3\bu_{j,h}^{n+1} - 4\bu_{j,h}^n + \bu_{j,h}^{n-1},
\bv_h
\right)
+ b^*(2\bu_{j,h}^{n}-\bu_{j,h}^{n-1}, \bu_{j,h}^{n+1}, \bv_h)
\nonumber\\- (p_{j,h}^{n+1}, \nabla \cdot \bv_h)
+  (\nu_j\nabla \bu_{j,h}^{n+1}, \nabla \bv_h)
&= (\bif(t_{n+1}), \bv_h),
\label{eq:FOM_momentum}
\\
(\nabla \cdot \bu_{j,h}^{n+1}, q_h) &= 0,
\label{eq:FOM_continuity}
\end{align}
 The nonlinear convection term is written in skew-symmetric form
\[b^*(\bu,\bv,\bw)
= \frac{1}{2}
(\bu \cdot \nabla \bv, \bw)
-\frac{1}{2}(\bu \cdot \nabla \bw, \bv).
\]

The discrete system \eqref{eq:FOM_momentum}–\eqref{eq:FOM_continuity} constitutes the high-fidelity FOM, which provides solutions accurate to $O(\Delta t^2 + h^{2m})$ for the $(\mathbb{P}_m, \mathbb{P}_{m-1})$ or $(\mathbb{P}_m, \mathbb{P}_{m-1}^{disc})$ stable elements \cite{garcia2023robust}. The resulting velocity snapshots $\{\bu_{1,h}^1,\bu_{1,h}^2,\cdots,\bu_{1,h}^{N},\bu_{2,h}^1,\bu_{2,h}^2,\cdots,\bu_{2,h}^{N},\cdots,\bu_{J,h}^1,\bu_{J,h}^2,\cdots,\bu_{J,h}^{N}\}$, are subsequently used to construct the ROM basis via POD. We also run FOM simulation for the viscosity $\E[\nu]$ using the following scheme:\\ Find $(\btu_h^{n+1},\tilde{p}_h^{n+1})\in  \bX_h \times Q_h,$ such that  $\forall (\bv_h, q_h) \in \bX_h \times Q_h$ \begin{align}
\left(
\frac{3\btu_{h}^{n+1} - 4\btu_{h}^n + \btu_{h}^{n-1}}{2\Delta t},
\bv_h
\right)
+ b^*(2\btu_{h}^{n}-\btu_{h}^{n-1}, \btu_{h}^{n+1}, \bv_h)
\nonumber\\- (\tilde{p}_{h}^{n+1}, \nabla \cdot \bv_h)
+  (\E[\nu]\nabla \btu_{h}^{n+1}, \nabla \bv_h)
&= (\bif(t_{n+1}), \bv_h),
\label{eq:FOM_momentum-expec}
\\
(\nabla \cdot \btu_{h}^{n+1}, q_h) &= 0.
\label{eq:FOM_continuity-expec}
\end{align}

\section{Stochastic Collocation Methods (SCMs) for UQ of fluid flows} \label{scm-uq} The uncertainty in the model often depends on high-dimensional random variables; consequently, the model suffers from the curse of dimensionality \cite{stoyanov2016dynamically}. Sampling based methods such as MCMs~\cite{rubinstein2016simulation} or ML-MCMs~\cite{giles2008multilevel} require a huge number of realizations, and thus quite demanding. In contrast, SCMs exploit the smooth dependence of the solution on the random parameters, require far fewer realizations to achieve the desired level of accuracy, and remain nonintrusive. Thus, SCMs can easily be combined with legacy codes without changing the PDE solver. In the popular sparse grid \cite{nobile2008sparse} SCM, the PDEs are solved for the solution $\bu(\bx, t_n, \by_j)$ at a given set of collocation points $\{ \by_j: \by_j\in \bGamma,\;j=1,2,\cdots, J \}$ and time $t_n$, then construct a global polynomial interpolation as $$\bu^{sc}(\bx,t_n,\by)=\sum_{l=1}^{N_p}c_l(\bx,t_n)\boldsymbol{\phi}_l(\by),$$ where $\{\phi_l\}_{l=1}^{N_p}$ is a basis of the finite-dimensional interpolating space $\mathcal{P}_{N_p}\subset \bL_\rho^2(\bGamma)$ and $c_l(\bx,t_n)$ is the coefficient of the basis $\boldsymbol{\phi}_l$. In SCMs, $J$ denotes the number of stochastic collocation points. For a Quantity of Interest (QoI) $\psi$ such as kinetic energy, SCMs approximate the multi-dimensional integral in the expectation value as below: \begin{align}
\mathbb{E}[\psi (\bu)] = \int_{\bGamma} \psi (\bu, \by)\rho (\by) \, d\by\approx
\sum_{j=1}^{J} w_j \psi (\bu, \by_j).\label{exp-scm}
\end{align}

SCMs have been employed extensively in UQ across diverse scientific and engineering applications, including the NSEs \cite{islam2026efficient,raveendran2025efficient}, magnetohydrodynamics (MHD) \cite{mohebujjaman2024efficient}, elliptic PDEs with random inputs \cite{babuska2007stochastic}, porous-media flow \cite{ganis2008stochastic}, mechanical and structural systems \cite{bressolette2010stochastic}, and heat-conduction problems involving uncertain parameters \cite{wang2016uncertainty}.

\section{Construction of G-ROM}\label{G-ROM-construction} For UQ of fluid flows, the standard Galerkin projection ROM (G-ROM) is obtained by representing the velocity field with the following POD expansion: 
\begin{equation}
\bu_r^j(\bx, t, \by_j) \equiv \sum_{k=1}^r a_k^j(t, \by_j) \bphi_k(\bx),
\end{equation}
where the unknown time-varying generalized coordinates $\{ a_k^j(t, \by_j) \}_{k=1}^r$ represent the POD-Galerkin trajectories. If the POD basis is weakly divergence-free, projecting the NSEs onto the reduced space gives the standard G-ROM. Specifically, for each $i=1,\ldots,r$ and $j=1,\ldots,J$, the reduced solution satisfies
\begin{align}
\biggl(
\frac{\partial \bu_r^j}{\partial t}, \bphi_i
\biggr)
+ 
\biggl(
\nu(\by_j)\nabla \bu_r^j,\nabla \bphi_i
\biggr)
+
\biggl(
(\bu_r^j\cdot\nabla)\bu_r^j,\bphi_i
\biggr)
=0.
\label{g-rom}
\end{align} 
This reduced finite-dimensional system may be written equivalently as a system of ODEs for the generalized coordinate vector $\ba_j(t)=(a^j_k)$: For $j=1,2,\cdots,J$,
\begin{equation}
\dot{\ba}_j = \A \ba_j + \ba_j^\top \B \ba_j,\label{g-rom-matrix}
\end{equation}
where the operators $\A$ and $\B$ correspond to the stiffness and nonlinear matrices in the numerical discretization of the NSEs \eqref{strong-form-problem}, respectively. That is, in ROM, we solve an initial value system of ODEs instead of solving the system of PDEs in \eqref{strong-form-problem}. In component-wise form, \eqref{g-rom-matrix} can be presented as below: For a fixed $j=1,2,\cdots,J$, for all $k = 1, \ldots, r$,
\begin{equation}
\dot{a}_k^j = \sum_{m=1}^r \A_{km}a_m^j + \sum_{m=1}^r \sum_{n=1}^r \B_{kmn}a_n^ja_m^j,
\end{equation}
where $\A_{km} = -(\nu(\bx, \by_j)\nabla \bphi_m, \nabla \bphi_k)$ and $\B_{kmn} = -
\bigl(
\bphi_m \cdot \nabla \bphi_n, \bphi_k
\bigr)$.

\subsection{ROM Basis}\label{rom-basis}
We employ POD to construct the ROM basis, although the proposed framework can also accommodate alternative data-driven basis-generation techniques, including Dynamic Mode Decomposition (DMD) \cite{schmid2022dynamic,tu2013dynamic}. Let $\{\bu_h^1,\bu_h^2,\ldots,\bu_h^N\}\subset \bX_h$ be snapshot solutions of \eqref{strong-form-problem} at $N$ time instances (or parameter realizations). While the snapshots in this work come from a FE discretization, any high-fidelity solver outcomes or empirical data could be used. We measure optimality in the $L^2$-norm, i.e., subject to the inner product
\[
(\bv,\bw)=\int_{\cD}\bv\cdot\bw\,\mathrm{d}\bx,
\]
whose FE counterpart is $(\bv,\bw)_{L^2}\approx \bv^\top M_h \bw$, where $M_h$ is the FE mass matrix. Given a target dimension $r\ll \mathscr{N}:=\dim \bX_h$, the POD seeks an $L^2$–orthonormal set $\{\bphi_k\}_{k=1}^r \subset \bX_h$ that minimizes the mean squared reconstruction error of the snapshots:
\begin{equation}\label{eq:POD-min}
\min_{\substack{\bphi_1,\ldots,\bphi_r\in X_h\\ (\bphi_i,\bphi_k)_{L^2}=\delta_{ik}}}
\;\sum_{n=1}^N \big\|\bu_h^n - \sum_{k=1}^r (\bu_h^n,\bphi_k)_{L^2}\,\bphi_k\big\|_{L^2}^2,
\end{equation}
where $\delta_{ik}$ is the Kronecker delta. The solution of \eqref{eq:POD-min} is characterized by an eigenvalue problem built from the snapshots. Let $\Y=[\,\bu_h^1,\; \bu_h^2,\; \cdots\;, \bu_h^N\,]\in\mathbb{R}^{\mathscr{N}\times N}$ collect the FE coefficient vectors of the snapshots, and define the  correlation matrix
\[
C := \Y^\top M_h \Y \;\in\; \mathbb{R}^{N\times N},
\]
which is symmetric positive semi-definite.
Compute eigenpairs $\{(\lambda_k,\bpsi_k)\}_{k=1}^N$ of $C$,
\begin{equation}\label{eq:snap-eig}
\Y^\top M_h \Y\, \bpsi_k = \lambda_k \bpsi_k,\qquad \bpsi_i^\top \bpsi_k=\delta_{ik}.
\end{equation}
Let $R:=\operatorname{rank}(M_h^{1/2}\Y)\le \min\{\mathscr{N},N\}$. The eigenvalues satisfy
\[
\lambda_1 \ge \lambda_2 \ge \cdots \ge \lambda_R > 0, \qquad \lambda_{R+1}=\cdots=\lambda_N=0.
\]
The POD modes (FE coefficient vectors) are then
\begin{equation}\label{eq:modes}
\bphi_k \;=\; \frac{1}{\sqrt{\lambda_k}}\, \Y\, \bpsi_k \;\in\; \mathbb{R}^{\mathscr{N}},\qquad k=1,\ldots,R,
\end{equation}
which are $L^2$–orthonormal in the discrete sense: $\bphi_i^\top M_h \bphi_k=\delta_{ik}$. Equivalently, if $M_h^{1/2}\Y=U\Sigma V^\top$ is an SVD, then $\lambda_k=\sigma_k^2$ and $\bphi_k=M_h^{-1/2}U_{:,k}$. For any $r\le R$, the $r$–dimensional POD space is
\begin{equation}
\bX_r := \operatorname{span}\{\bphi_1,\ldots,\bphi_r\}.
\end{equation}
The POD provides the best $r$–dimensional $L^2$ approximation of the snapshots: the optimal value of \eqref{eq:POD-min} equals the {discarded energy}
\[
\sum_{n=1}^N \big\|\bu_h^n - \Pi_r \bu_h^n\big\|_{L^2}^2 \;=\; \sum_{k=r+1}^{R} \lambda_k,
\]
where $\Pi_r$ is the $L^2$–orthogonal projector onto $\bX_r$. A common criterion for selecting $r$ is to prescribe a tolerance $\tilde{\varepsilon}\in(0,1)$ and choose the smallest $r$ such that
\[
\frac{\sum\limits_{k=1}^{r}\lambda_k}{\sum\limits_{k=1}^{R}\lambda_k} \;\ge\; 1-\tilde{\varepsilon}.
\]
If each snapshot satisfies the homogeneous boundary condition and the discrete incompressibility constraint, then so do all POD modes (being linear combinations of the snapshots). When pressure is included in the reduced formulation, one may enforce stability via standard techniques (e.g., supremizer enrichment); this does not affect the construction above. The same construction applies if the snapshots are drawn across multiple parameter (e.g., viscosity) realizations; the resulting basis captures the dominant coherent structures averaged over those realizations.

\section{Numerical Experiments}\label{numerical-exp}

 \begin{figure}[h!]
    \centering
    \includegraphics[width=0.5\textwidth]{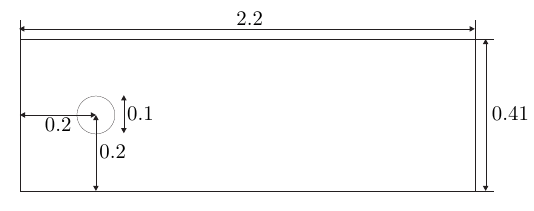} 
    \caption{Geometry of the channel-flow configuration with a circular cylinder.}
    \label{fig:channel}
\end{figure}

This section evaluates the proposed methods for the stochastic NSEs \eqref{strong-form-problem} using the two-dimensional flow past a circular cylinder as the benchmark problem. Consider $\bx=(x_1,x_2)$. The computational domain consists of a $2.2 \times 0.41$ rectangular channel containing a circular cylinder of radius $0.05$, whose center is located at $(0.2,\,0.2)$; see Fig.~\ref{fig:channel}. No-slip boundary conditions are imposed on the channel walls as well as on the surface of the cylinder. Prescribed inflow and outflow velocity profiles are adopted in accordance with established benchmark configurations \cite{john2004reference,rebholz2017improved}, namely,
\begin{align}
    u_1(0,x_2,t) &= u_1(2.2,x_2,t) = \frac{6}{0.41^2} x_2(0.41-x_2),\label{bdcon-1}\\
u_2(0,x_2,t) &= u_2(2.2,x_2,t) = 0.\label{bdcon-2}
\end{align}

No external body force is applied $(\bif=\textbf{0})$ and we consider the expected kinematic viscosities $\E[\nu]=10^{-3}$ and $\E[\nu]=2\times10^{-4}$, corresponding to random Reynolds numbers with mean values $Re=100$ and $Re=500$, respectively. These parameter choices correspond to moderate-Reynolds-number regimes commonly used in ROM studies \cite{noack2011reduced}. In the stochastic simulations presented later, these expected viscosities serve as the nominal values about which the random viscosity realizations are constructed. For the realization-independent nonhomogeneous boundary data, the snapshot mean is subtracted before constructing the ROM basis, so that the resulting fluctuation snapshots and POD modes satisfy homogeneous boundary conditions.

\begin{algorithm}
	\caption{Data-Driven Correction Stochastic Collocation ROM (DDC-SCM-ROM)}\label{ERF-SCM-ROM-algo}
	\begin{algorithmic}[1]
		\Procedure \red{OFFLINE} construction 
		\State \textbf{Initialization:} Mesh, FE functions, $T_{train}$, $N_{train}$, $N_{sc}$, $\bu^{0}_h$, snapshot matrices: $\Y=[\hspace{1mm}]$, $\widetilde{\Y}=[\hspace{1mm}]$; $\{\by_{j}\}_{j=1}^{N_{train}}$, $ \{\nu_{j}\}_{j=1}^{N_{train}}$, Tolerance: $tol_d$, $tol_c$, $\epsilon$; $N'$, $k^*$, $M$
		\For{$n = 0, \cdots, N'$}
		\For{$j = 1, \cdots, N_{train}$}
		\State Compute $\bu^{n+1}_{j,h}$ by solving \eqref{eq:FOM_momentum}-\eqref{eq:FOM_continuity} for viscosity $\nu_j$
        \If{$n\ge k^*$}
        \State Store: $\Y = [\Y, \bu^{n+1}_{j,h}]$
        \EndIf
		\EndFor
		\EndFor
		\State Solve for the reduced basis $\{\bphi_k\}_{k=1}^{r}$  according to \eqref{eq:modes}
        \State Compute $\A,\B$ 
        \For{$n = 0, \cdots, N'$}
        \State Compute $\btu_h^{n+1}$ by solving \eqref{eq:FOM_momentum-expec}-\eqref{eq:FOM_continuity-expec} for viscosity $\E[\nu]$
        \If{$n\ge k^*$}
        \State Store: $\widetilde{\Y} = [\widetilde{\Y}, \btu_h^{n+1}]$
        \EndIf
        \EndFor
        \State Compute $ \widetilde{\A}$ and $\widetilde{\B}$ by solving \eqref{time-uncos} using $tol_d$ and snapshots $\widetilde{\Y}$
        \State Compute $\widehat{\A},$ and $\widehat{\B}$ by solving \eqref{const-optimi-data} using $tol_c$, $\epsilon$ and snapshots $\widetilde{\Y}$
		\EndProcedure
		\Procedure \red{ONLINE} Construction
		\State \textbf{Initialization:} $T$, $N_{sc}$, $\bu_{r}^{0}$,
		$\{\by_{j}\}_{j=1}^{N_{sc}}$, $\{w_{j}\}_{j=1}^{N_{sc}}$, $ \{\nu_{j}\}_{j=1}^{N_{sc}}$
		\For{$n = 0, \cdots, N$}
		\For{$j = 1, \cdots, N_{sc}$}
		\State Find $\bu_{r,j}^{n+1}$:
        \If {DDC-UQ-ROM}
        \State Solve \eqref{eqn-1.9}
        \ElsIf {CDDC-UQ-ROM}
        \State Solve \eqref{eqn-1.9-cDDC-UQ-ROM}
        \EndIf
		\State Calculate $\psi(\bu_{r,j}^{n+1})$
		\EndFor
		\State Estimate $\mathbb{E}[\psi(\bu^{n+1})] \approx \sum\limits_{j = 1}^{N_{sc}} w_{j} \psi(\bu_{r,j}^{n+1})$
		\EndFor
		\EndProcedure
	\end{algorithmic}
\end{algorithm}

\subsection{Variable 5D Random Viscosity}

In this section, we approximate the random viscosity by $\nu(\bx, \by)$, where $\by = (y_1, y_2, \ldots, y_5) \in \Gamma \subset \mathbb{R}^5$ is a 5D random vector. As the SCM, we employ the Clenshaw–Curtis sparse grid \cite{stoyanov2016dynamically} where the quadrature weights and  the high dimensional stochastic collocation points are generated by Tasmanian \cite{stoyanov2015tasmanian}. The mean and covariance of the viscosity field are given by
\[
\mathbb{E}[\nu](\bx) = \frac{c}{\zeta}, \quad \text{Cov}[\nu](\bx, \bx') = \frac{1}{\zeta^2} \exp\left(-\frac{(\bx - \bx')^2}{\ell^2} \right),
\]
where $\ell$ is the correlation length. For the random Reynolds number with mean $Re=100$, we set $\zeta=1000$, whereas for the random Reynolds number with mean $Re=500$, we set $\zeta=5000$. The viscosity field is represented using a Karhunen--Lo\'eve expansion \cite{levy2008karhunen}:
\begin{align}
\nu(\bx, \by_j(\omega)) = \frac{1}{\zeta} \bigg(
c + \sqrt{\frac{\pi \ell^2}{2}} y_1^j(\omega) + \sum_{k=1}^{q} \sqrt{\xi_k} \Big[ & \sin\left(\frac{k \pi x_1}{2.2} \right) \sin\left(\frac{k \pi x_2}{0.41} \right) y_{2k}^j(\omega) \nonumber \\
& + \cos\left(\frac{k \pi x_1}{2.2} \right) \cos\left(\frac{k \pi x_2}{0.41} \right) y_{2k+1}^j(\omega) \Big] \bigg),
\label{eq:KL_expansion}
\end{align}
where the infinite series is truncated after the first $q$ terms. The random variables $y_k^j$ are uncorrelated with zero mean and unit variance, and the eigenvalues are given by
\[
\sqrt{\xi_k} = \sqrt{\pi \ell} \exp\left(-\frac{(k \pi \ell)^2}{8} \right).
\]
For our test case, we take $\by_j(\omega)=(y_1^j(\omega),y_2^j(\omega),\cdots,y_d^j(\omega)) \in [-\sqrt{3}, \sqrt{3}]^d$, $\ell = 0.01$, $d = 5$, $c = 1$, and $q = 2$.

\subsection{In-Sample and Out-of-Sample ROM}

For the in-sample parameter ROM, both the offline and online ROMs correspond to the 11 different viscosities associated with the $N_{\text{train}}=N_{sc}=11$ collocation points. On the other hand, for the out-of-sample ROM, the offline ROM corresponds to the $N_{\text{train}}=11$ collocation points, while the online ROM corresponds to 801 different viscosities associated with $N_{sc}=801$ collocation points in the 5D parameter space.

\subsection{Direct Numerical Simulation (DNS)} To evaluate the ROM performance, we use the DNS kinetic energy as a benchmark for the G-ROM, DDC-UQ-ROM, and CDDC-UQ-ROM. Based on the 5D parameter space, we generate two sets of sparse-grid stochastic collocation points and their corresponding SCM quadrature weights. One set contains 11 and the other set contains 801 stochastic collocation points. Through the Karhunen--Lo\`eve expansion in \eqref{eq:KL_expansion}, we generate two sets of viscosities corresponding to these two sets of the collocation points. For the in-sample ROMs, the benchmark DNS kinetic energy is computed using \(N_{\mathrm{sc}}=11\) five-dimensional collocation points and their corresponding weights, whereas for the out-of-sample ROMs, it is computed using \(N_{\mathrm{sc}}=801\) points and their corresponding weights.

In both cases, we use the pointwise divergence-free \((\mathbb{P}_2,\mathbb{P}_1^{disc})\) Scott--Vogelius (SV) element \cite{scott1985norm} on a barycenter-refined triangular mesh, yielding 35,020 velocity and 25,974 pressure degrees of freedom (DoFs). 

For each viscosity, we first solve a stationary Stokes problem subject to the boundary conditions in \eqref{bdcon-1}--\eqref{bdcon-2}. Its solution is then used as the initial condition for a linearized backward-Euler (BE) fully discrete scheme at the first time step. The Stokes and BE solutions provide the two initial conditions required for the BDF-2 scheme, subject to the same boundary conditions.

For each viscosity, the FOM in \eqref{eq:FOM_momentum}-\eqref{eq:FOM_continuity} is solved over \(t\in[0,17]\) with \(\Delta t=0.002\). That is, $N=8500$. At each time step, the estimated kinetic energy is computed using the weighted average in \eqref{exp-scm}. For these two different DNS, the solutions at \(t=7\) are stored for use as initial conditions in the online in-sample and out-of-sample ROMs. That is, $k^*=3500$ in Algorithm \ref{ERF-SCM-ROM-algo}.

We perform the DNS for  $\ell=0.01$ and $0.1$ and then estimate the expected kinetic energy using the following SCM weighted average
\[
\mathbb{E}[\psi(\bu_h^{n})]
\approx
\sum_{j=1}^{N_{\mathrm{sc}}} w_j
\psi(\bu_{j,h}^{n})
\]
as the primary quantity of interest. In this case, $\psi(\bu_{j,h}^{n})=\frac{1}{2}\|\bu_h(\bx,t_n,\by_j)\|^2$.

\subsection{Offline ROM} 

To construct the ROM basis, ROM matrices (mass, stiffness, and nonlinear matrices), and the data-driven operators $\widetilde{\A}$, $\widetilde{\B}$, $\widehat{\A}$ and $\widehat{\B}$ we generate the snapshots using the FOM. To this end, we employ the $(\mathbb{P}_2,\mathbb{P}_1^{disc})$ SV element on a barycenter-refined triangular mesh, which provides exactly the same DoFs as in the DNS. Similar to the DNS, we generate the two initial conditions required for the second-order time-stepping schemes. We then run the BDF-2 scheme to obtain solutions at subsequent time steps, subject to the boundary conditions in \eqref{bdcon-1}--\eqref{bdcon-2}, until we reach $T=17$ with $\Delta t=0.002$. Snapshots are collected at each time step from $t=7$ to $t=10$.

\subsubsection{Generation of Ensemble Snapshots and UQ ROM Basis for Random Reynolds Number with Mean $Re=100$ and $500$}
\begin{figure}
	\centering
	\subfloat[]{\includegraphics[width=0.4\textwidth,height=0.25\textwidth]{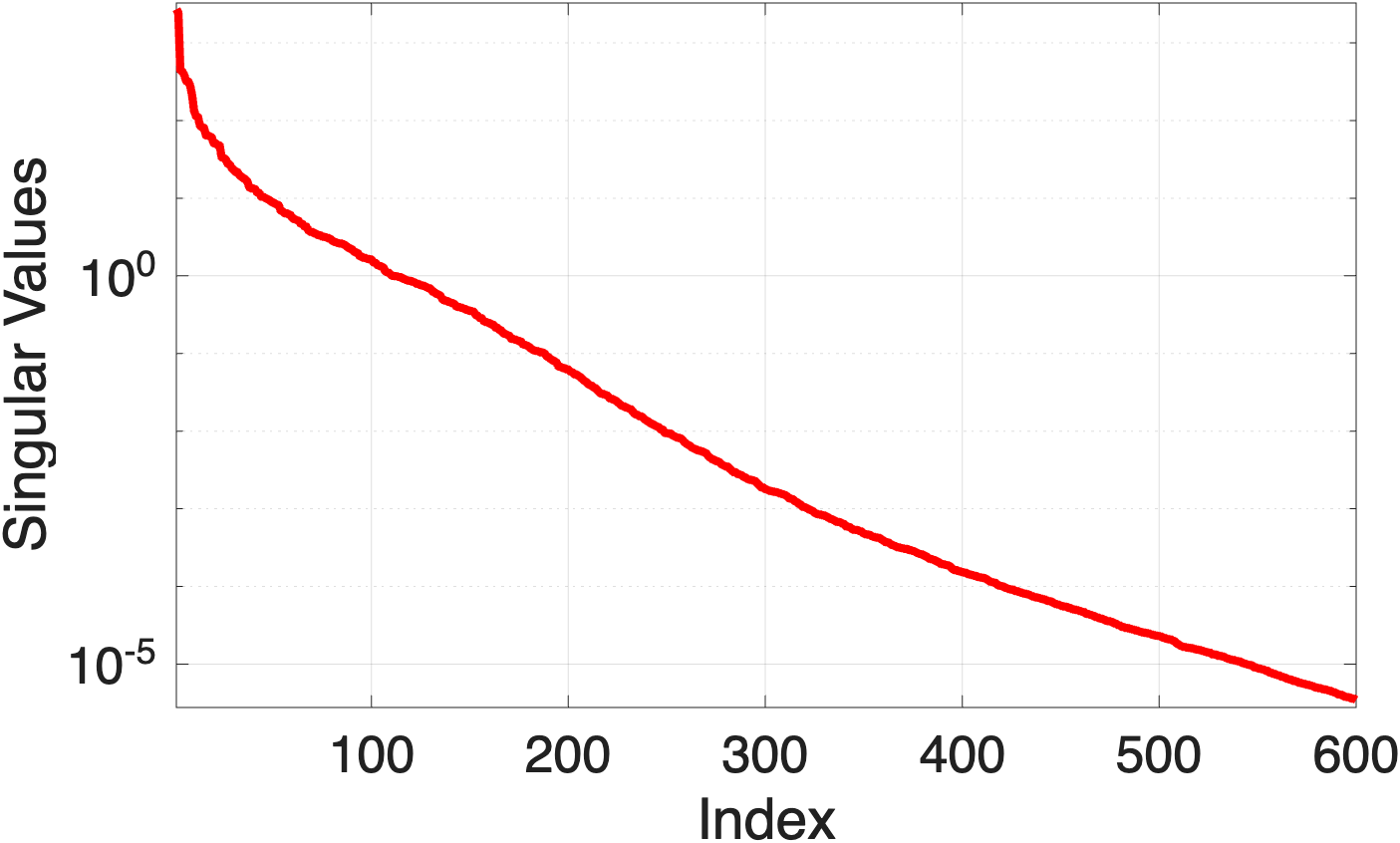}}\hspace{3ex}
	\subfloat[]{\includegraphics[width=0.4\textwidth,height=0.25\textwidth]{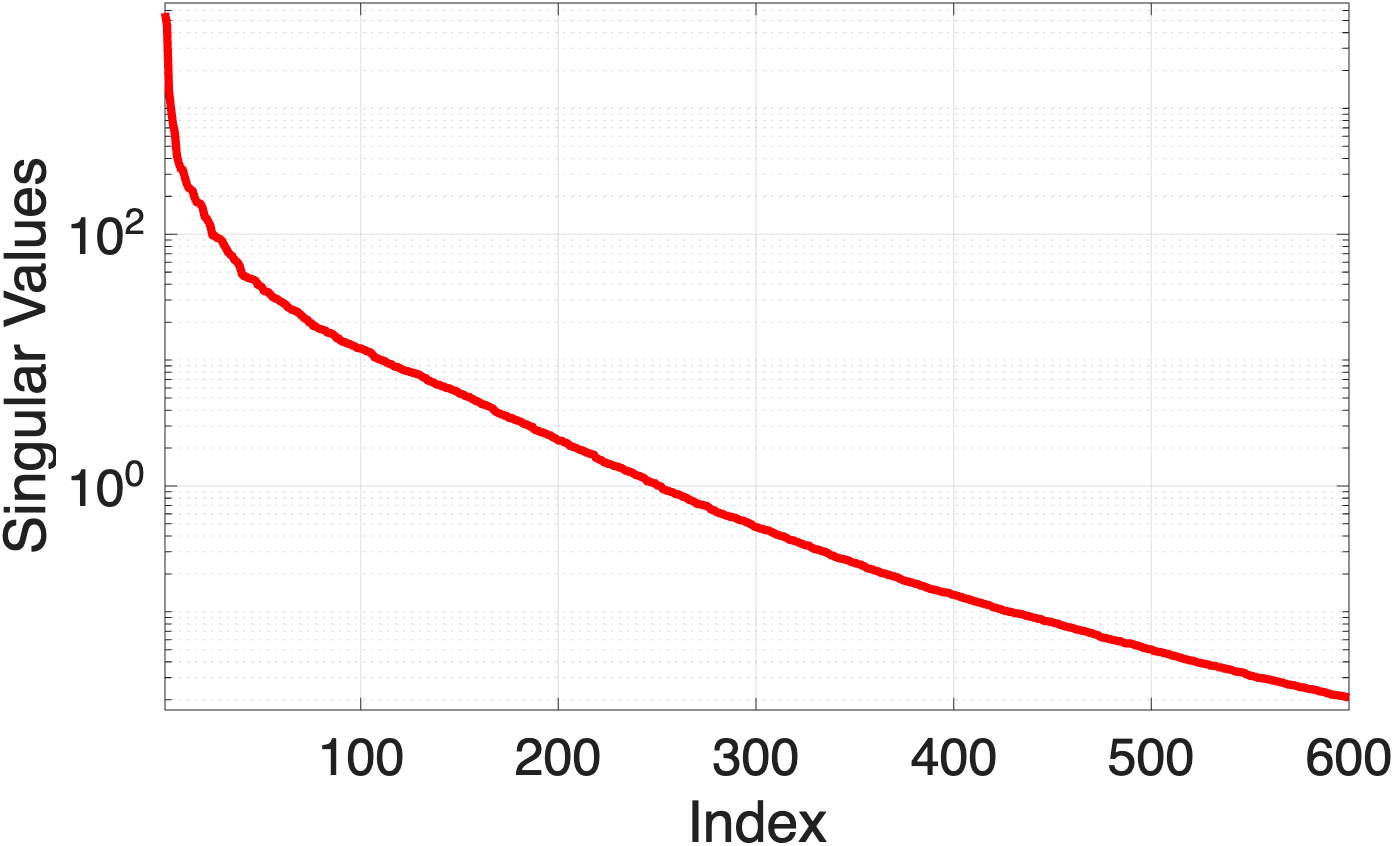}}\vspace{-4ex}
	\caption{Decay of the singular values for the flow past a cylinder problem on a semilogarithmic scale for  random Reynolds number with mean (a) $Re=100$ and (b) $Re=500$.}\label{singular-values-plot}\vspace{-4ex}
\end{figure}

For the UQ ROM basis generation, we generate 
$\{\by_j\}_{j=1}^{N_{\text{train}}}$ sparse-grid collocation points and their corresponding weights in the 5D parameter space using Tasmanian, with $N_{\text{train}}=11$. For these collocation points, we generate 11 corresponding viscosities through the Karhunen--Lo\`eve expansion \eqref{eq:KL_expansion}. FOM simulations using \eqref{eq:FOM_momentum}-\eqref{eq:FOM_continuity} are performed independently for each of these viscosities, and snapshots are collected at each time step from $t=7$ to $t=10$. We then construct the snapshot matrix $\Y$ as follows:
\[
\Y=[\bu_S^{1,k^*},\bu_S^{1,k^*+1},\cdots,\bu_S^{1,N^{'}},
\bu_S^{2,k^*},\bu_S^{2,k^*+1},\cdots,\bu_S^{2,N^{'}},
\cdots,
\bu_S^{N_{\text{train}},k^*},\bu_S^{N_{\text{train}},k^*+1},
\cdots,\bu_S^{N_{\text{train}},N^{'}}],
\]
where $\bu_S^{j,k}$ is the snapshot of the $j^{\text{th}}$ realization at time $t_k=(k^*+k)\Delta t$, for $0\leq k\leq N^{'}$, and $N^{'}=1500$. The first mode is selected as the average of all snapshots, which is then subtracted from each column of $\Y$. We then solve the eigenvalue problem in \eqref{eq:snap-eig} using this modified snapshot matrix to obtain the solution of the optimization problem given in \eqref{eq:POD-min}. The singular values are plotted in Fig.~\ref{singular-values-plot} for random Reynolds number with mean (a) $Re=100$ and (b) $Re=500$.

\subsubsection{Generation of the Data-Driven ROM Operators $\widetilde{\A}$, $\widetilde{\B}$, $\widehat{\A}$ and $\widehat{\B}$}

In the FOM simulation using \eqref{eq:FOM_momentum-expec}--\eqref{eq:FOM_continuity-expec} with viscosity $\mathbb{E}[\nu]=10^{-3}$, the flow reaches a periodic-in-time (statistically steady) state at approximately $t=5$ following an initial spin-up period \cite{mohebujjaman2017energy}. This is the case for the random Reynolds number with mean $Re=100$. Thus, we form the snapshot matrix $\widetilde{\Y}$ using the snapshots collected at each time step from $t=7.002$ to $t=7.332$, that is,
\[
\widetilde{\Y}=[\btu_S^{k^*+1},\btu_S^{k^*+2},\cdots,\btu_S^{k^*+M}],
\]
where $\btu_S^{k}$ represents the snapshot of the FOM with viscosity $\mathbb{E}[\nu]=10^{-3}$ at $t_k=(k^*+k)\Delta t$, for $1\le k\le M$, $k\in\mathbb{Z}$, $k^*=3500$ and $M=166$. These 166 snapshots constitute one complete period \cite{mohebujjaman2019physically}. We compute the subfilter-scale stress tensor, $\btau_{\Bar{\nu}}$, in \eqref{sub-filter-scale-stress-tensor} with $d=16$ for each of these snapshots. Then, using these snapshots, we solve the unconstrained optimization problem in \eqref{time-uncos} with the user-defined tolerance $tol_d$ to obtain $\widetilde{\A}$ and $\widetilde{\B}$, and the constrained optimization problem in \eqref{const-optimi-data} with the given tolerance $tol_c$ and $\epsilon$ to obtain $\widehat{\A}$ and $\widehat{\B}$.

For the random Reynolds number with mean $Re=500$, the data-driven ROMs operators are constructed using a separate training window consisting of $M=219$ consecutive snapshots spanning the interval $[10.002,10.438]$. These snapshots constitute one complete period \cite{mohebujjaman2019physically}. In this case, we set $k^*=5000$. As in the case with mean $Re=100$, we solve \eqref{time-uncos} and \eqref{const-optimi-data} to obtain the data-driven correction ROM operators

\subsection{Online ROM}

The online ROM simulations are performed using the dominant modes constructed during the offline ROM phase. These modes capture the spatial structure and therefore satisfy the prescribed boundary conditions. Recall that, during the FOM (DNS) simulations, the FE solutions at $t=7$ were stored. We project these solutions onto the ROM space $\bX_r$ in the $L^2$ sense and use the projected solutions as the initial conditions for the online ROM simulations. As in the DNS, taking the solutions at $t=7$ as the initial state (which is relabeled as $t=0$ for the online ROM), we compute the ROM solutions for all viscosity samples at the first time step using a linearized BE scheme.

For the remaining time steps, we run the G-ROM using the following BDF-2 discretization of \eqref{g-rom}: for all $i=1,\ldots,r$ and $n=1,2,\ldots$, compute $\bu_{j,r}^{n+1}$ such that
\begin{align}
    \frac{1}{2\Delta t}\left(
    3\bu_{j,r}^{n+1}-4\bu_{j,r}^{n}+\bu_{j,r}^{n-1},
    \bphi_i
    \right)
    +\left(\nu(\by_j)\nabla\bu_{j,r}^{n+1},\nabla\bphi_i\right)
    +b^*\left(
    2\bu_{j,r}^{n}-\bu_{j,r}^{n-1},
    \bu_{j,r}^{n+1},
    \bphi_i
    \right)
    =0,
\end{align}
for $j=1,2,\ldots,N_{\mathrm{sc}}$.

This yields the dynamical system in \eqref{g-rom-matrix} for the G-ROM. For the data-driven ROMs, using the operators $\widetilde{\A}$, $\widetilde{\B}$, $\widehat{\A}$, and $\widehat{\B}$ constructed during the offline phase, we obtain the dynamical systems in \eqref{eqn-1.9} and \eqref{eqn-1.9-cDDC-UQ-ROM} for the DDC-UQ-ROM and CDDC-UQ-ROM, respectively.

 In all tests, as in the FOM simulations, we use \(\Delta t=0.002\) and continue the online ROM simulations until \(t=10\), corresponding to \(t=17\) in the DNS simulations. For the in-sample and out-of-sample ROMs, we use \(N_{\mathrm{sc}}=11\) and \(N_{\mathrm{sc}}=801\), respectively.

We run the online ROM for several $r$ values and $\ell$ and then estimate the expected kinetic energy using the following SCM weighted average
\[
\mathbb{E}[\psi(\bu_r^{n})]
\approx
\sum_{j=1}^{N_{\mathrm{sc}}} w_j
\psi(\bu_{j,r}^{n}),
\]
where $\psi(\bu_{j,r}^{n})=\frac{1}{2}\|\bu_r(\bx,t_n,\by_j)\|^2$.

\subsubsection{Correlation length $\ell=0.01$} \label{cor-len-0-01} 
\begin{figure}
	\centering
	\subfloat[]{\includegraphics[width=0.45\textwidth,height=0.3\textwidth]{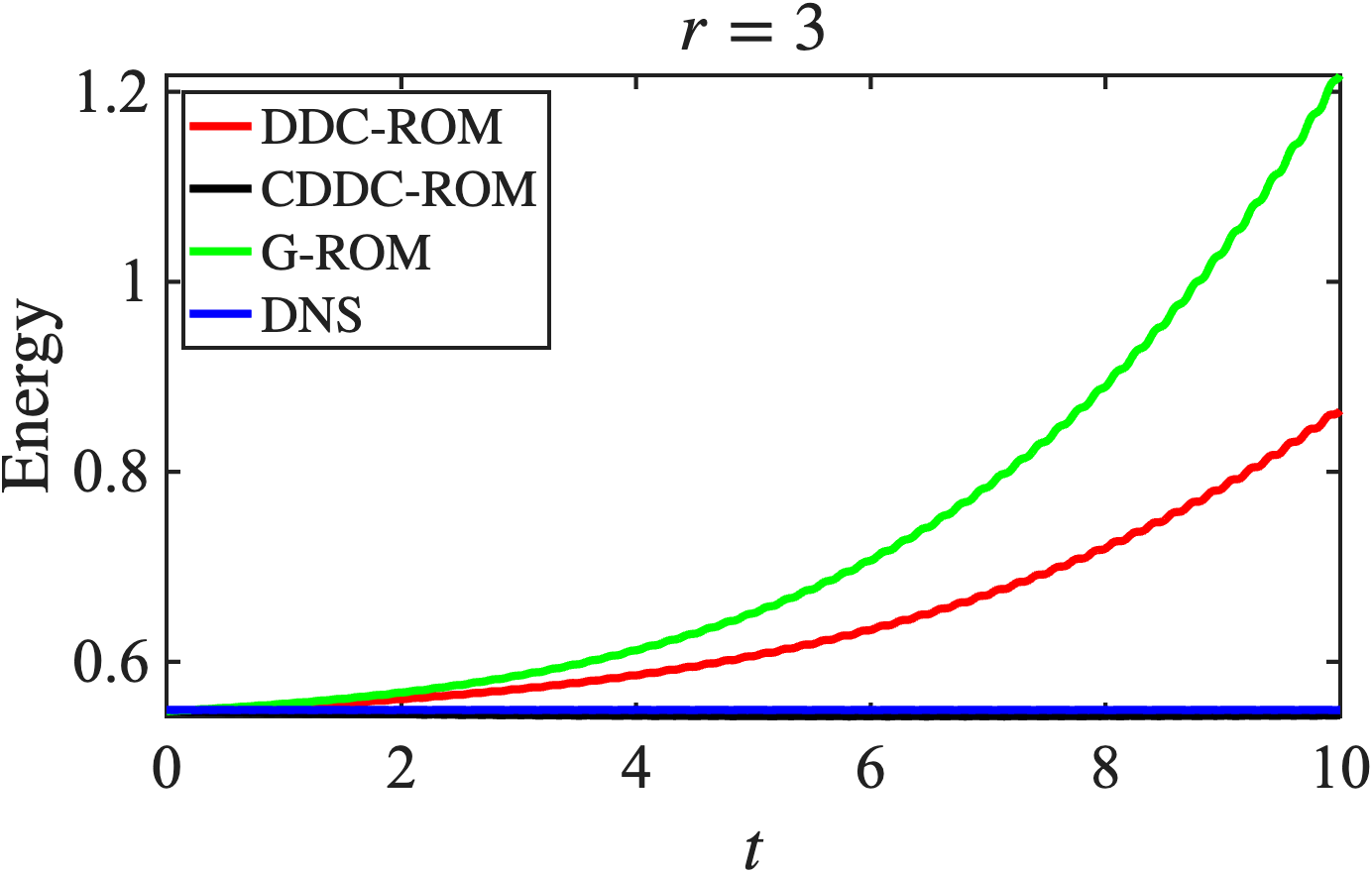}}\hspace{3ex}
	\subfloat[]{\includegraphics[width=0.45\textwidth,height=0.3\textwidth]{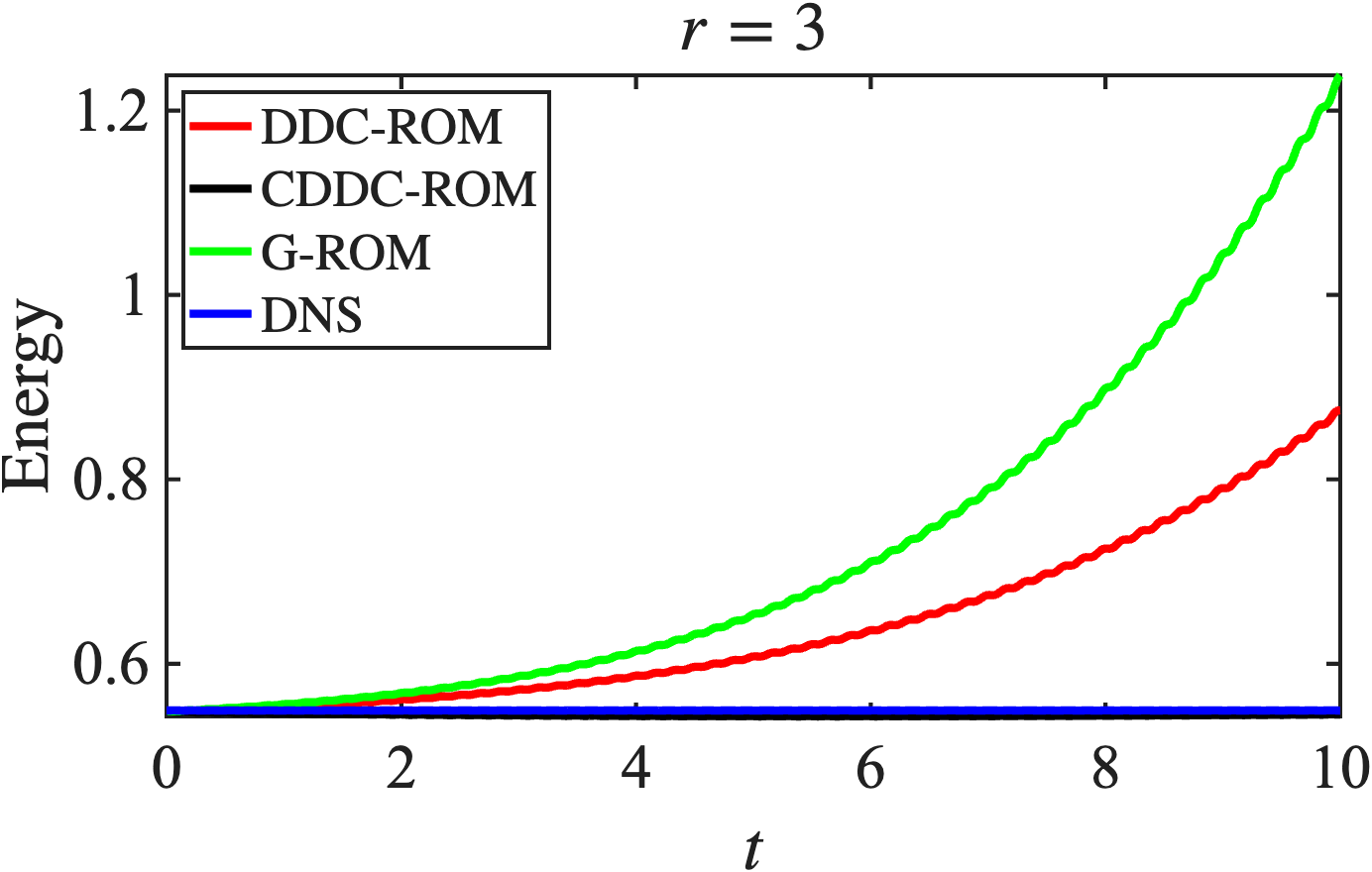}}\vspace{-4ex}
	\caption{Energy evolution for $r=3$ comparing DNS, G-ROM, DDC-UQ-ROM, and CDDC-UQ-ROM with $\ell = 0.01$, $tol_d=9.0\times 10^{-3}$, $tol_c=9.0\times 10^{-3}$, and $\epsilon= 6\times 10^{-2}$: (a) In-sample ROM, (b) Out-of-sample ROM for random Reynolds number with mean $Re=100$.}\vspace{-4ex}\label{r=3_T=10_wno_801}
\end{figure}

\begin{figure}
	\centering
	\subfloat[]{\includegraphics[width=0.45\textwidth,height=0.3\textwidth]{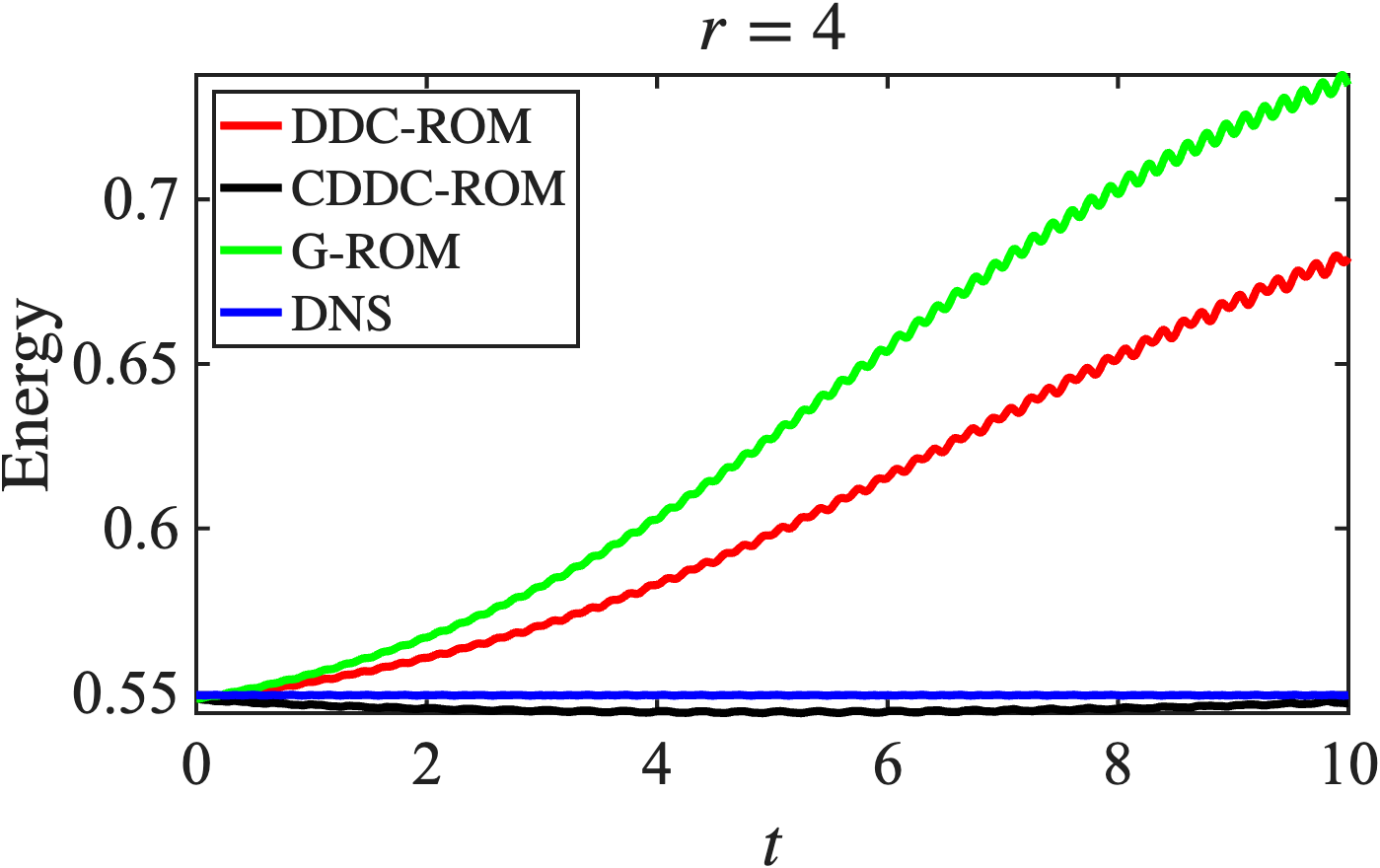}}\hspace{3ex}
	\subfloat[]{\includegraphics[width=0.45\textwidth,height=0.3\textwidth]{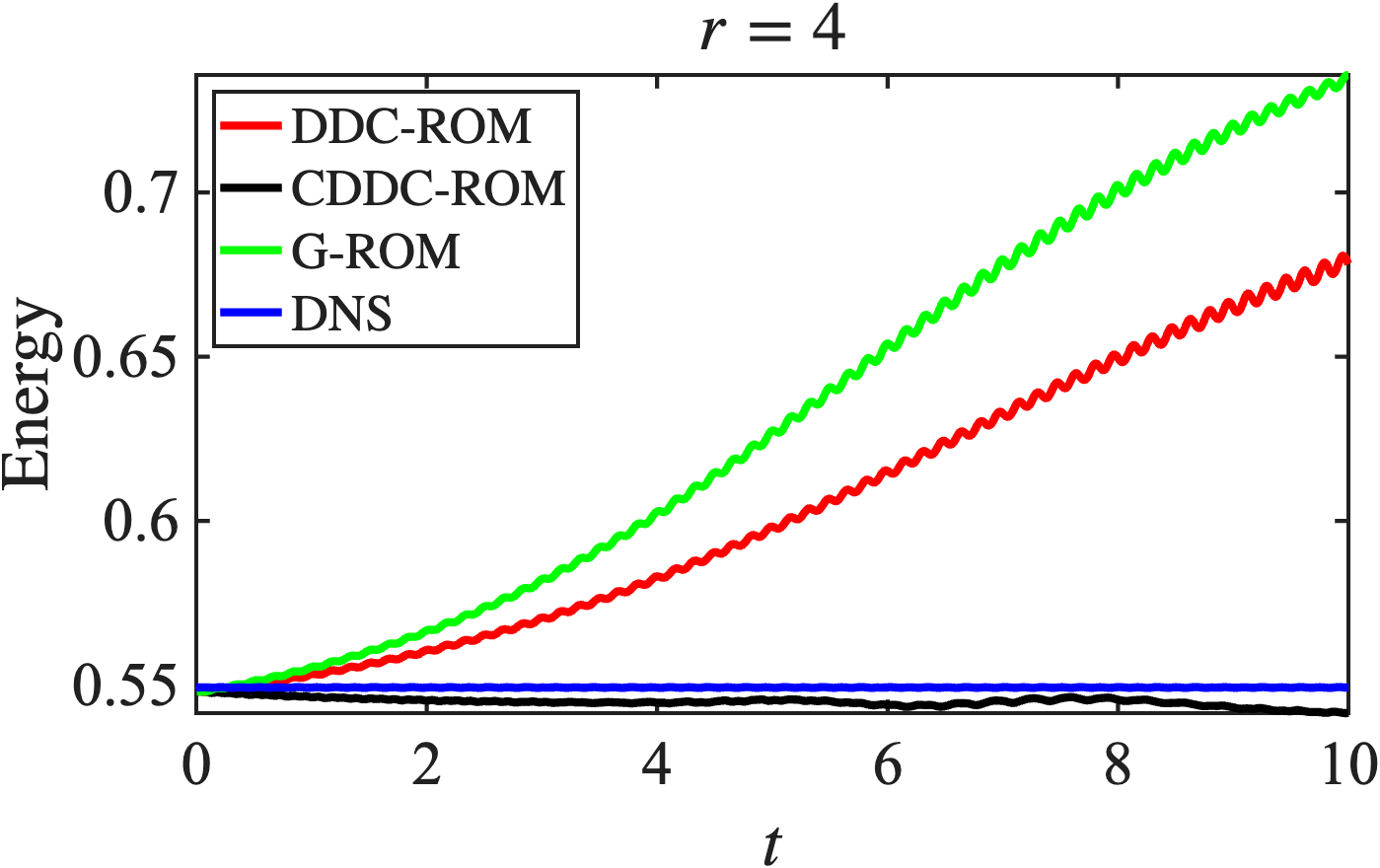}}\vspace{-4ex}
	\caption{Energy evolution for $r=4$ comparing DNS, G-ROM, DDC-UQ-ROM, and CDDC-UQ-ROM with $\ell = 0.01$, $tol_d=4.0\times 10^{-3}$, $tol_c=7.0\times 10^{-2}$, and $\epsilon= 3.5\times 10^{-2}$: (a) In-sample ROM, (b) Out-of-sample ROM  for random Reynolds number with mean $Re=100$.}\vspace{-4ex} \label{r=4_T=10_wno_801}
\end{figure}

\begin{figure}
	\centering
	\subfloat[]{\includegraphics[width=0.45\textwidth,height=0.3\textwidth]{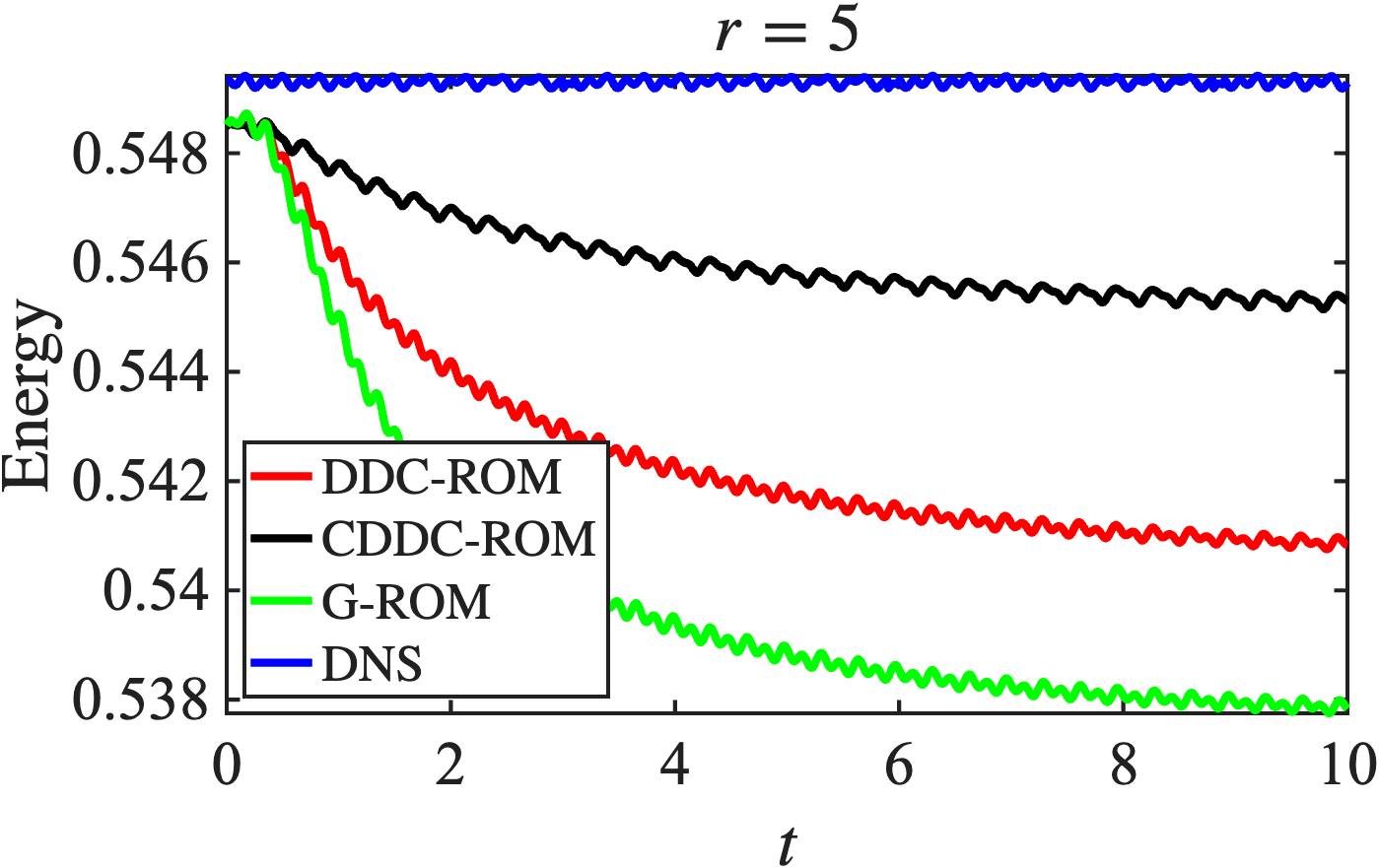}}\hspace{3ex}
	\subfloat[]{\includegraphics[width=0.45\textwidth,height=0.3\textwidth]{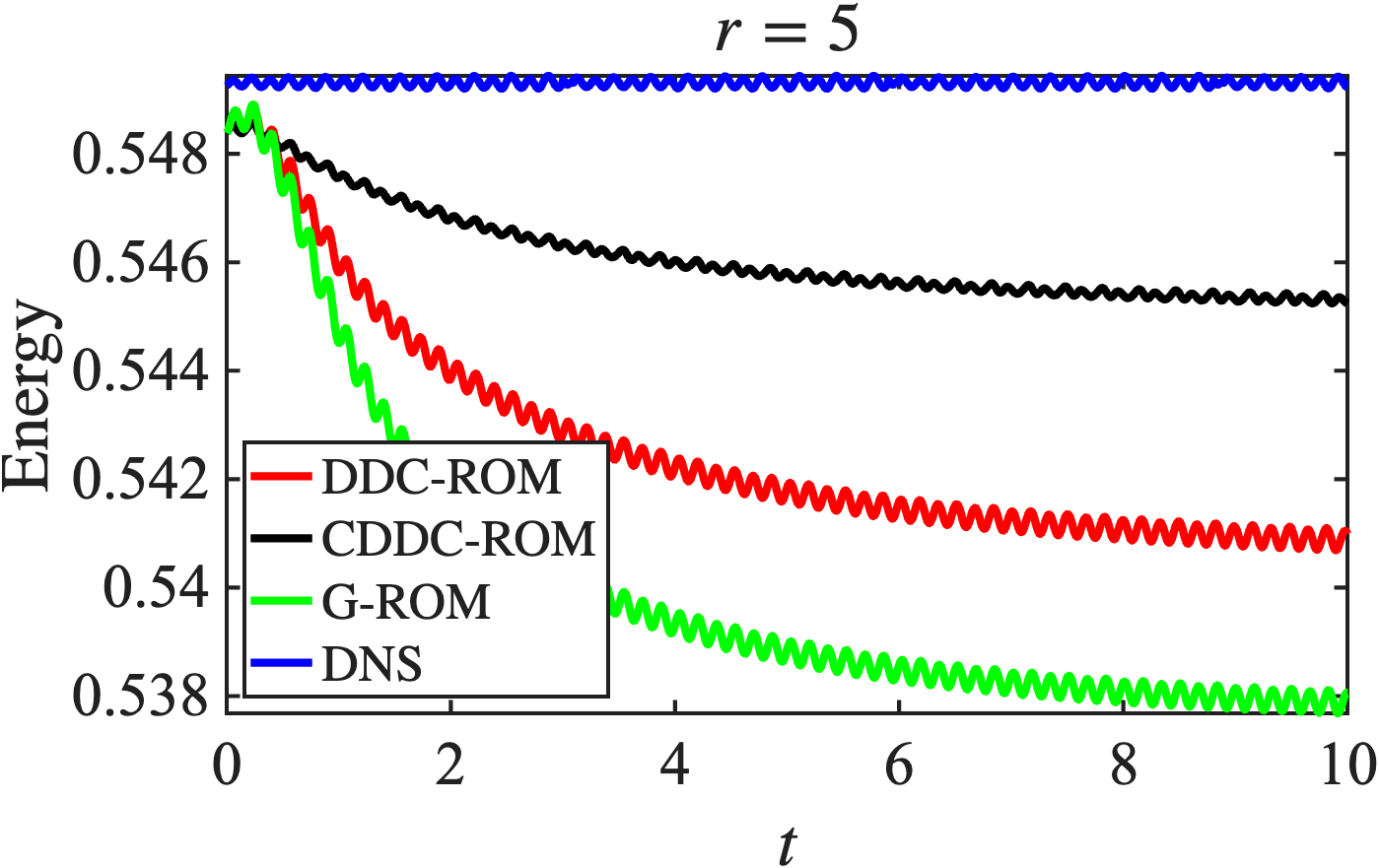}}\vspace{-4ex}
	\caption{Energy evolution for $r=5$ comparing DNS, G-ROM, DDC-UQ-ROM, and CDDC-UQ-ROM with $\ell = 0.01$, $tol_d=2.0\times 10^{-3}$, $tol_c=3.0\times 10^{-2}$, and $\epsilon= 2.4\times 10^{-2}$: (a) In-sample ROM, (b) Out-of-sample ROM  for random Reynolds number with mean $Re=100$.}\vspace{-4ex}	\label{r=5_T=10_wno_801}
\end{figure}

\begin{figure}
	\centering
	\subfloat[]{\includegraphics[width=0.45\textwidth,height=0.3\textwidth]{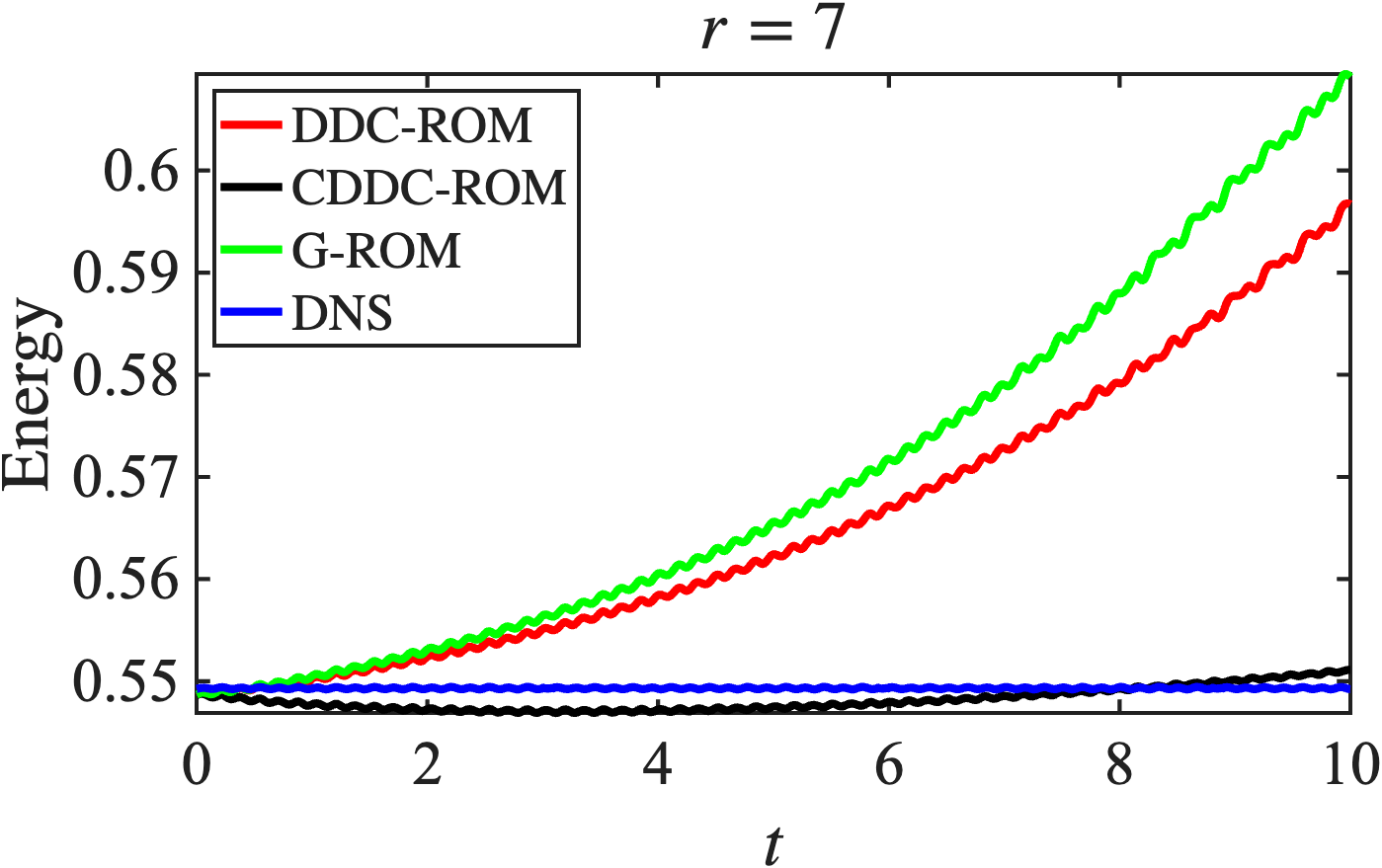}}\hspace{3ex}
	\subfloat[]{\includegraphics[width=0.45\textwidth,height=0.3\textwidth]{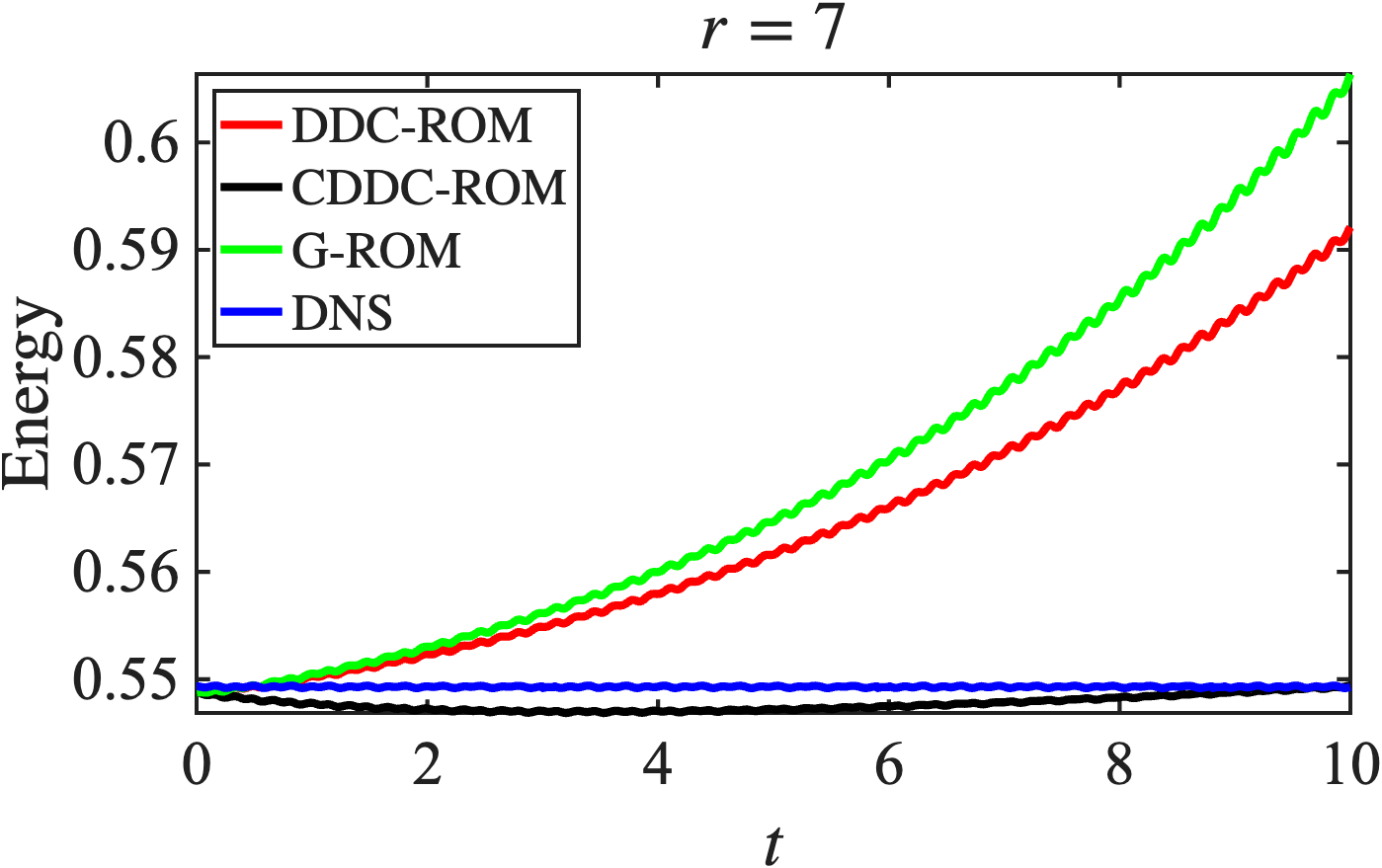}}\vspace{-4ex}
	\caption{Energy evolution for $r=7$ comparing DNS, G-ROM, CDDC-UQ-ROM, and DDC-UQ-ROM with $\ell = 0.01$, $tol_d=1\times 10^{-4}$, $tol_c=8\times 10^{-3}$, and $\epsilon= 2.2\times 10^{-4}$: (a) In-sample ROM, and (b) Out-of-sample ROM  for random Reynolds number with mean $Re=100$.}\vspace{-4ex}	\label{r=7_T=10_wno_801}
\end{figure}

\begin{figure}
	\centering
	\subfloat[]{\includegraphics[width=0.45\textwidth,height=0.3\textwidth]{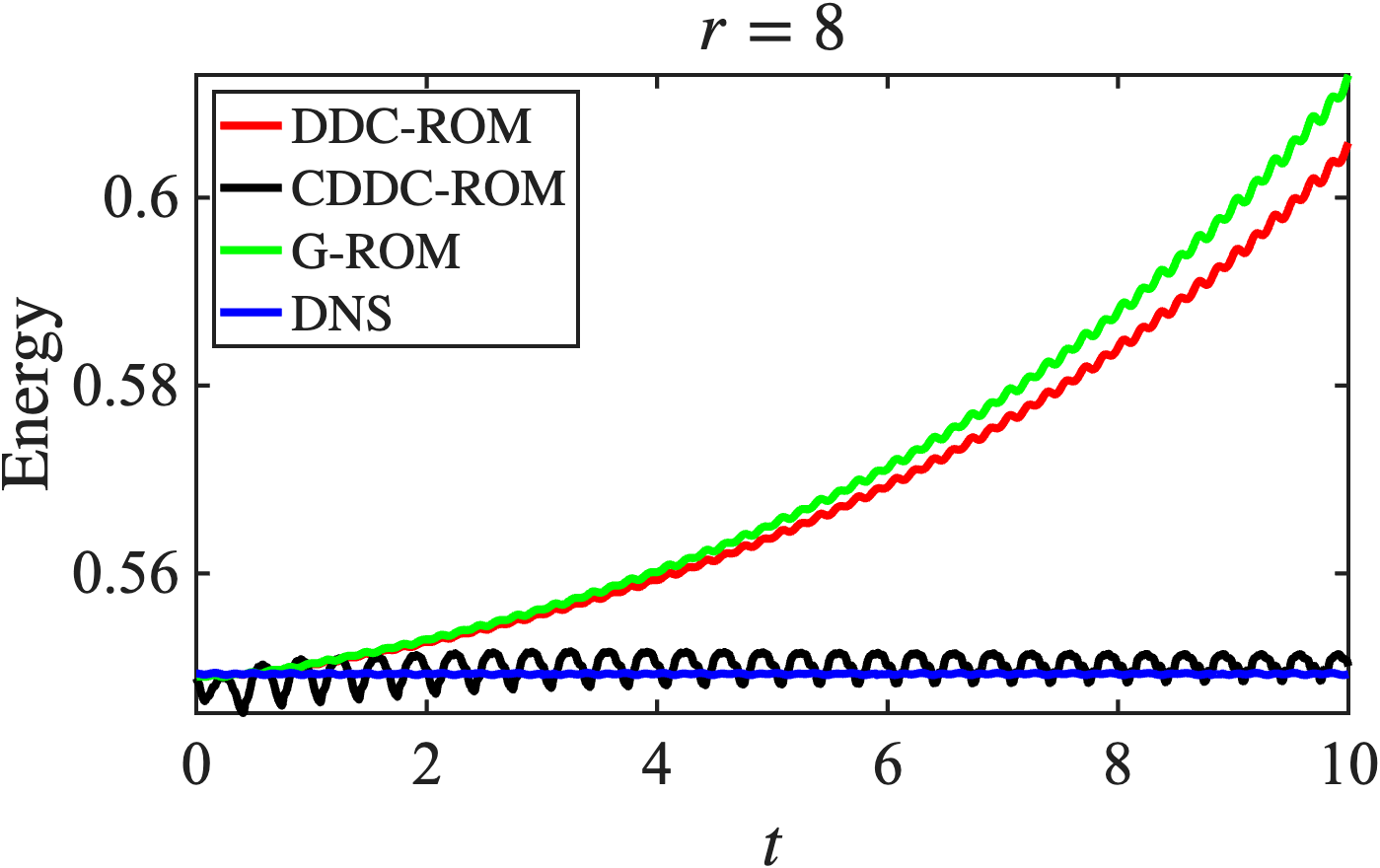}}\hspace{3ex}
	\subfloat[]{\includegraphics[width=0.455\textwidth,height=0.3\textwidth]{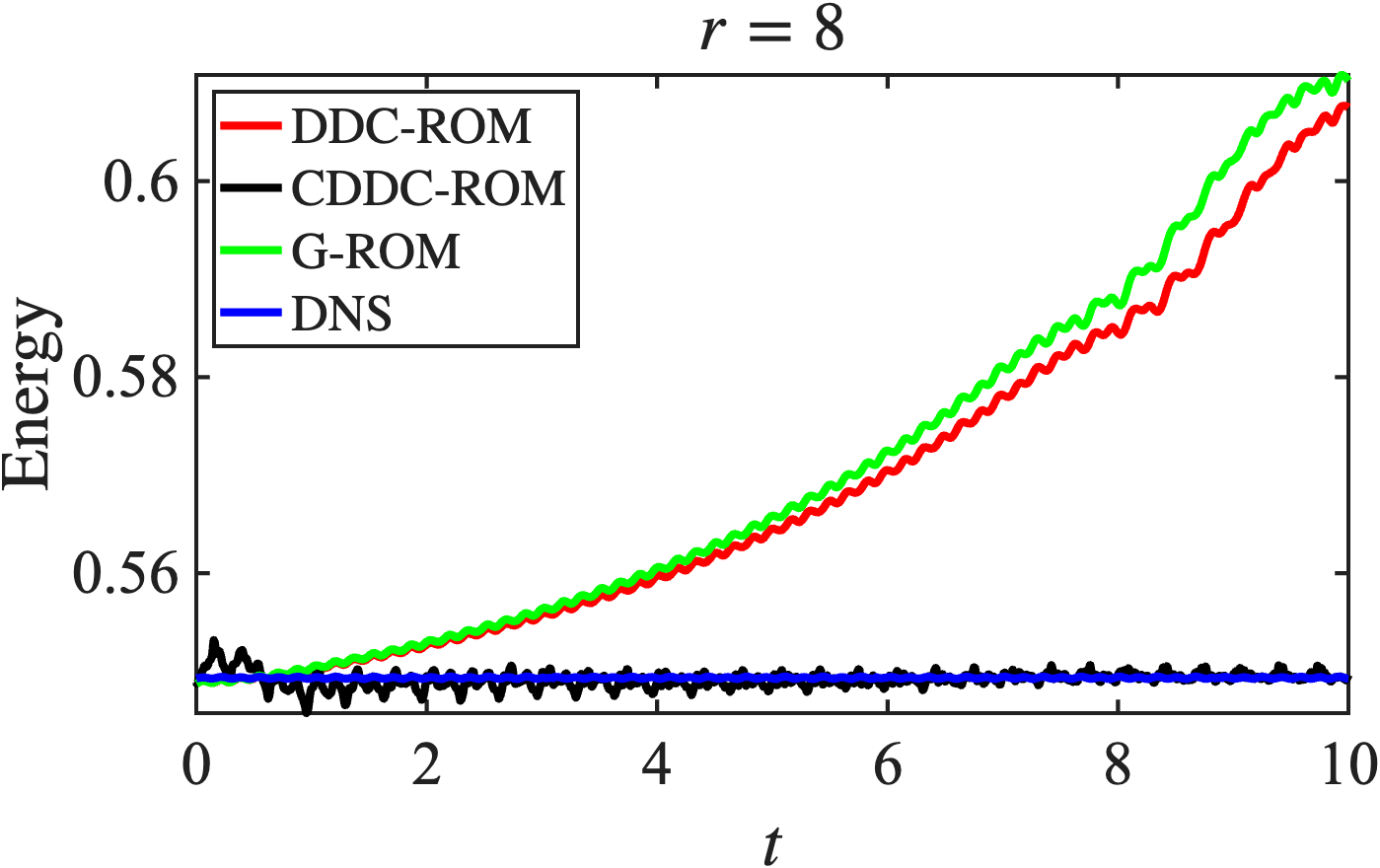}}\vspace{-4ex}
	\caption{Energy evolution for $r=8$ comparing DNS, G-ROM, CDDC-UQ-ROM, and DDC-UQ-ROM with $\ell = 0.01$, $tol_d=4\times 10^{-5}$, $tol_c=5\times 10^{-3}$, and $\epsilon= 3\times 10^{-10}$: (a) In-sample ROM, and (b) Out-of-sample ROM  for random Reynolds number with mean $Re=100$.} \vspace{-4ex}	\label{r=8_T=10_wno_801}
\end{figure}

This section evaluates the performance of the proposed DDC-UQ-ROM and CDDC-UQ-ROM for the stochastic NSE by comparing their results with those of the DNS and G-ROM. In Fig.~\ref{r=3_T=10_wno_801}--\ref{r=8_T=10_wno_801}, we set the correlation length to $\ell=0.01$ and random Reynolds number with mean $Re=100$. We plot the time evolution of the mean kinetic energy for the in-sample and out-of-sample parameter cases using different numbers of modes.

For \(r=3\), to determine the optimal data-driven operators, we use \(tol_d=9\times10^{-3}\) for the DDC-UQ-ROM and \(tol_c=9\times10^{-3}\) with \(\epsilon=6\times10^{-2}\) for the CDDC-UQ-ROM. Fig. \ref{r=3_T=10_wno_801} presents the time evolution of the mean kinetic energy for the (a) in-sample and (b) out-of-sample parameters for $r=3$. Both plots show that the data-driven ROMs are significantly more accurate than the G-ROM, with the CDDC-UQ-ROM providing the most accurate results. 

The G-ROM in Fig. \ref{r=4_T=10_wno_801} shows improved accuracy as the number of modes is increased to $r=4$ together with $tol_d=4.0\times 10^{-3}$, $tol_c=7.0\times 10^{-2}$ and $\epsilon=3.5\times 10^{-2}$. However, as in the case of $r=3$, the data-driven ROMs remain more accurate, with the CDDC-UQ-ROM yielding the most accurate results. 

In Fig. \ref{r=5_T=10_wno_801}, we further increase the number of modes to $r=5$ together with the data-driven ROM parameters $tol_d=2.0\times 10^{-3}$, $tol_c=3.0\times 10^{-2}$ and $\epsilon=2.4\times 10^{-2}$. We observe a significant improvement in the performance of both the G-ROM and DDC-UQ-ROM. The values of $r$ we have investigated, the DDC-UQ-ROM and G-ROM exhibit their best performance for $r=5$. Though, the CDDC-UQ-ROM demonstrates better performance than the DDC-UQ-ROM, while the DDC-UQ-ROM performs better than the G-ROM.

In Fig. \ref{r=7_T=10_wno_801} and Fig. \ref{r=8_T=10_wno_801}, we present the time evolution of the mean kinetic energy for $r=7$ and $r=8$, respectively. In Fig. \ref{r=7_T=10_wno_801}, we used $tol_d=1.0\times 10^{-4}$, $tol_c=8.0\times 10^{-3}$ and $\epsilon=2.2\times 10^{-4}$, whereas in Fig. \ref{r=8_T=10_wno_801}, we used $tol_d=4.0\times 10^{-5}$, $tol_c=5.0\times 10^{-3}$ and $\epsilon=3.0\times 10^{-10}$. We observe that, as the number of modes is increased beyond the optimal value $r=5$, the performance of both the G-ROM and DDC-UQ-ROM deteriorates. However, the DDC-UQ-ROM performs relatively better than the G-ROM. In contrast, the CDDC-UQ-ROM continues to exhibit superior performance and accurately captures the DNS results.

\begin{figure}
    \centering
    \subfloat[]{\includegraphics[width=0.45\textwidth,height=0.3\textwidth]{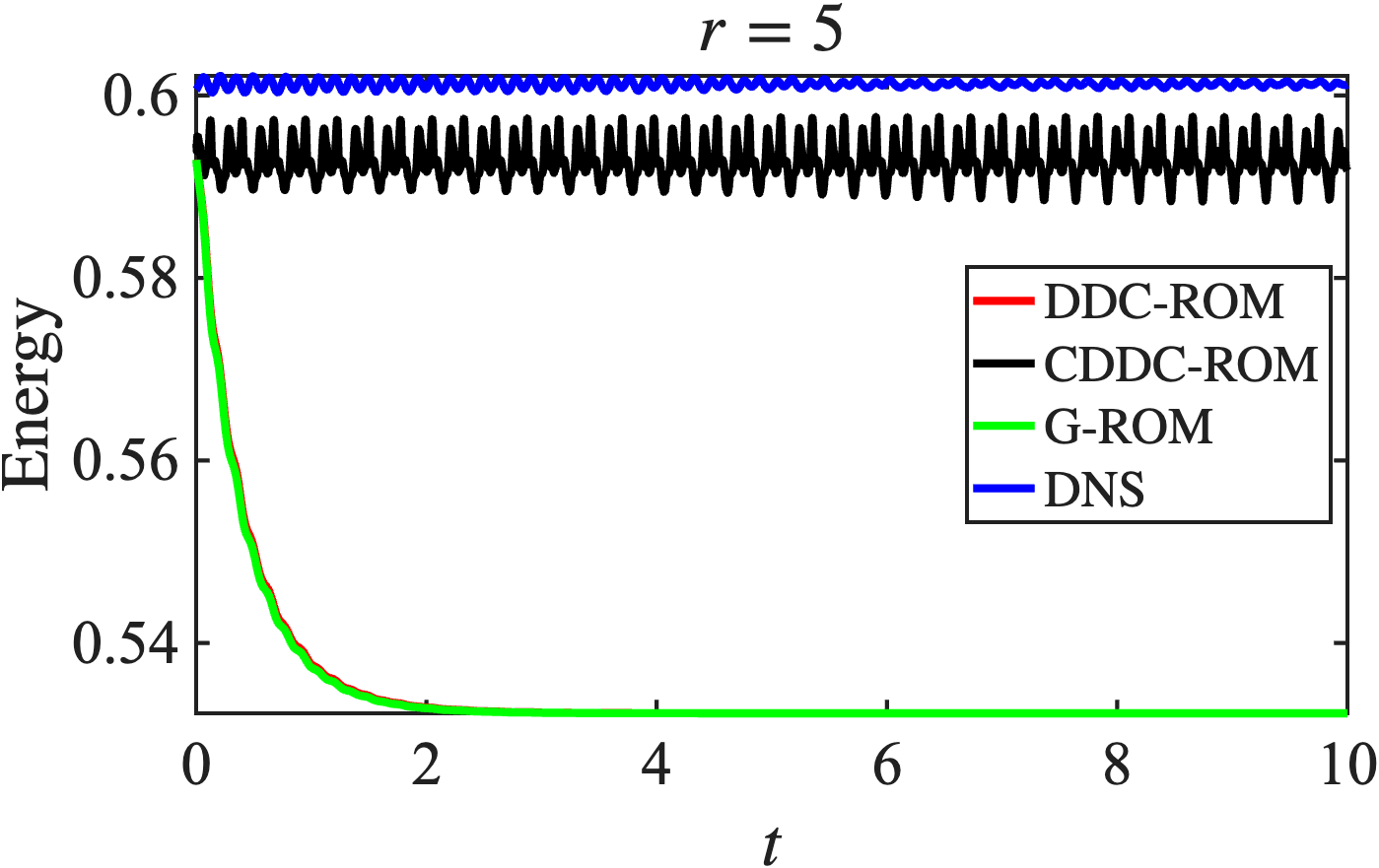}} \hspace{3ex}
    \subfloat[]{\includegraphics[width=0.45\textwidth,height=0.3\textwidth]{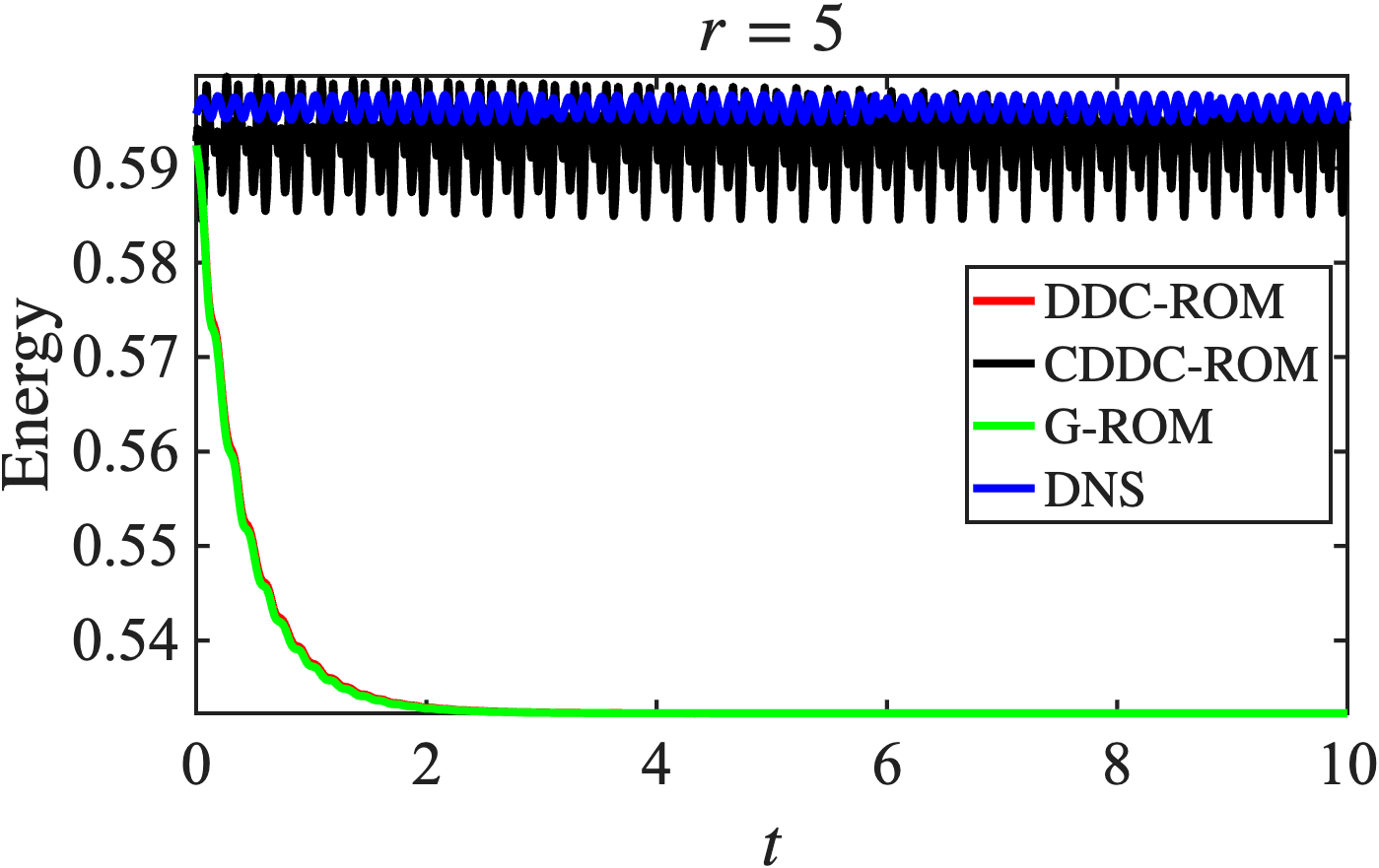}}\vspace{-4ex}

    \caption{Energy evolution for $r=5$ comparing DNS, G-ROM, CDDC-UQ-ROM, and DDC-UQ-ROM with $\ell = 0.01$, $tol_d=4\times 10^{-2}$, $tol_c=3\times 10^{-1}$, and $\epsilon= 3.5\times 10^{-2}$: (a) In-sample ROM, and (b) Out-of-sample ROM  for random Reynolds number with mean $Re=500$.}\vspace{-4ex}
    \label{fig:energy_Re500}
\end{figure}

The in-sample and out-of-sample ROM energy evolution over the time interval $[0,10]$ for $r=5$ and a random Reynolds number with mean $Re=500$ is shown in Fig. \ref{fig:energy_Re500}, comparing the DNS, G-ROM, CDDC-UQ-ROM, and DDC-UQ-ROM. We consider $tol_d=4\times 10^{-2}$, $tol_c=3\times 10^{-1}$, and $\epsilon= 3.5\times 10^{-2}$ to compute data-driven ROM operators $\widetilde{\A}$, $\widetilde{\B}$, $\widehat{\A}$ and $\widehat{\B}$. We observe that the G-ROM and DDC-UQ-ROM exhibit comparable accuracy, whereas the CDDC-UQ-ROM provides the most accurate approximation of the DNS in both the in-sample and out-of-sample cases for random Reynolds number with mean $Re=500$.

Overall, across the cases investigated, the CDDC-UQ-ROM significantly outperforms all other ROMs. These results demonstrate that the data-driven ROMs developed for the UQ problem generally outperform the G-ROM.

\subsubsection{Correlation length $\ell=0.1$}
\begin{figure}
	\centering
	\subfloat[]{\includegraphics[width=0.45\textwidth,height=0.3\textwidth]{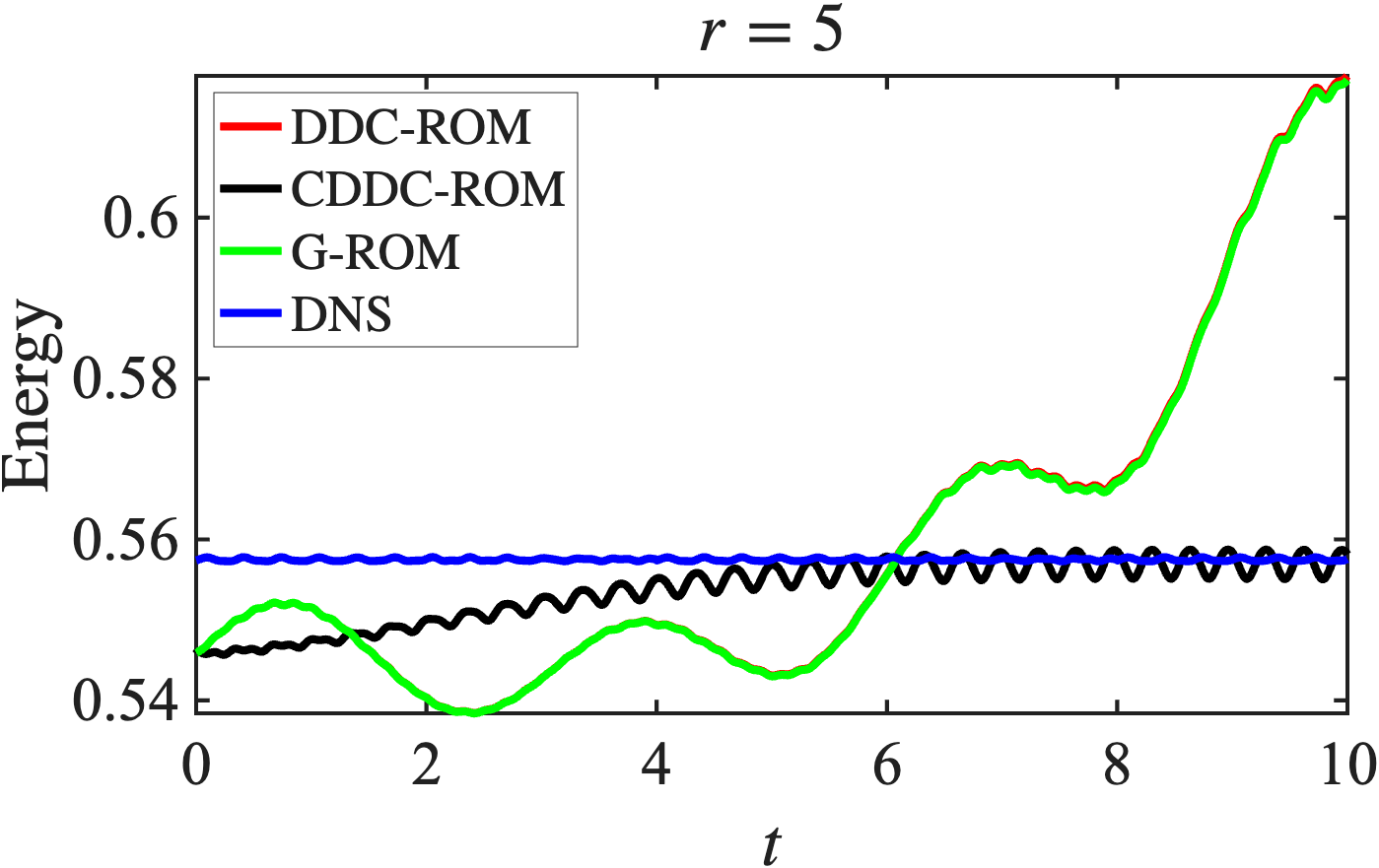}}\hspace{3ex}
	\subfloat[]{\includegraphics[width=0.45\textwidth,height=0.3\textwidth]{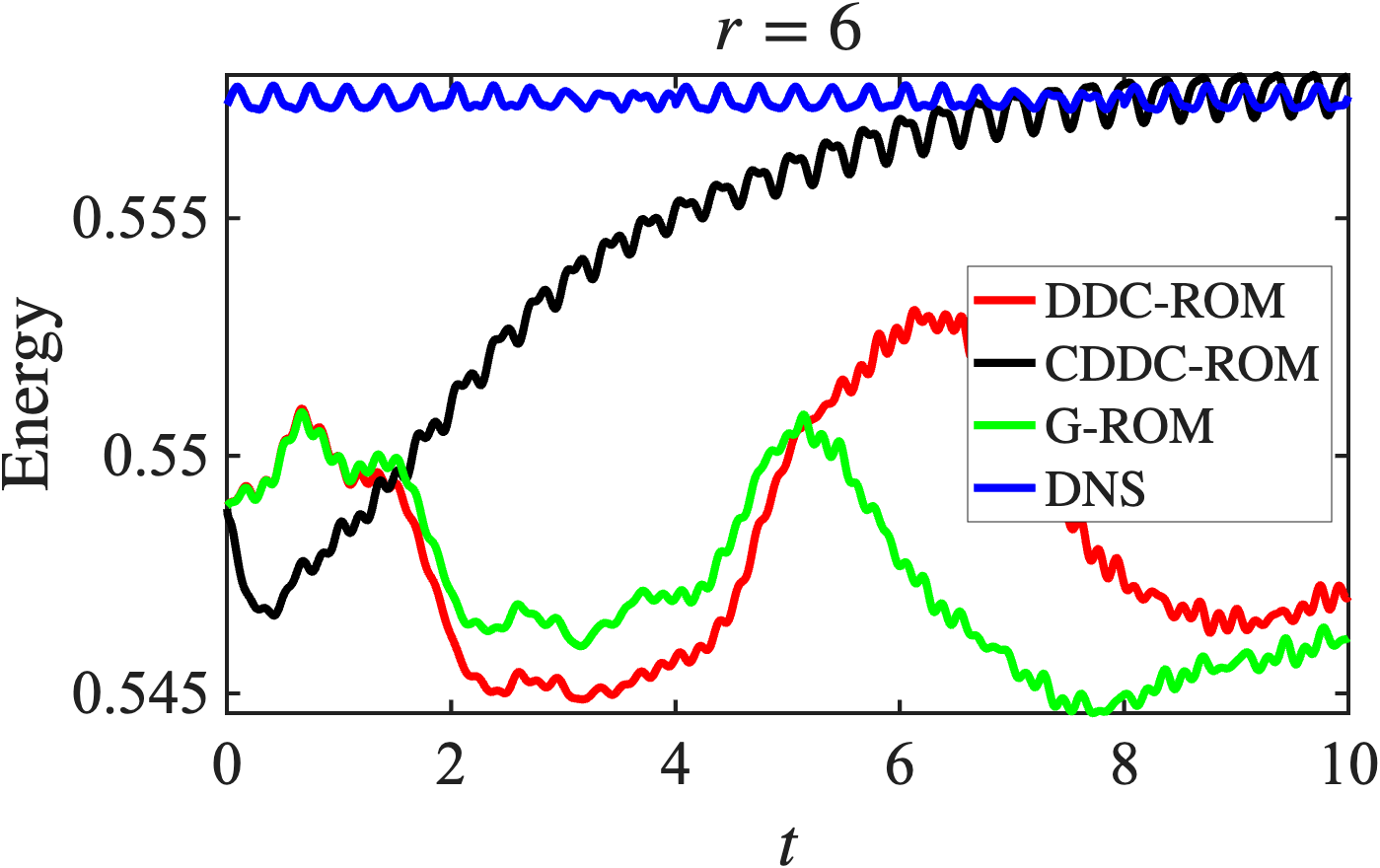}}\\\subfloat[]{\includegraphics[width=0.45\textwidth,height=0.3\textwidth]{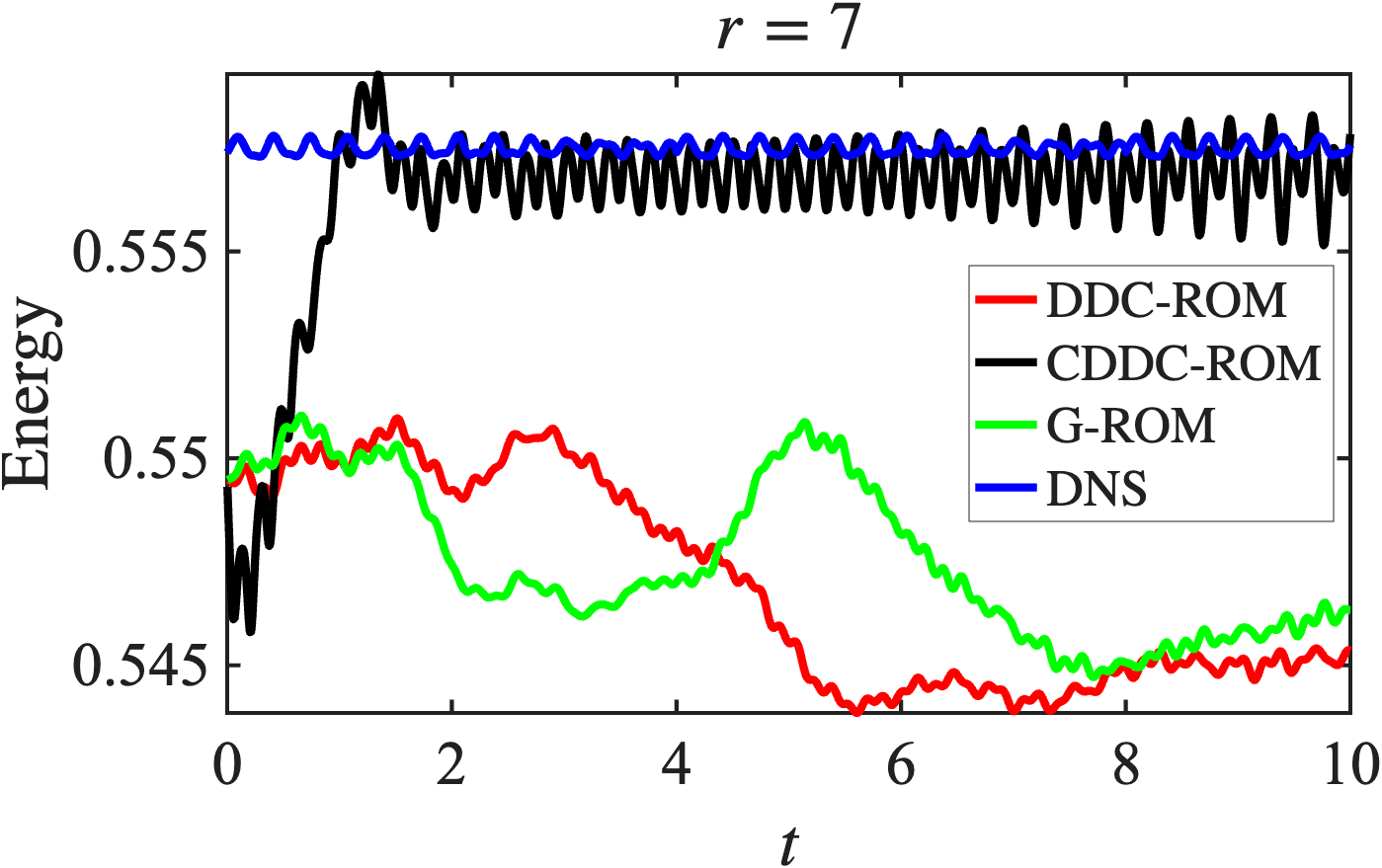}}\hspace{3ex}
	\subfloat[]{\includegraphics[width=0.45\textwidth,height=0.3\textwidth]{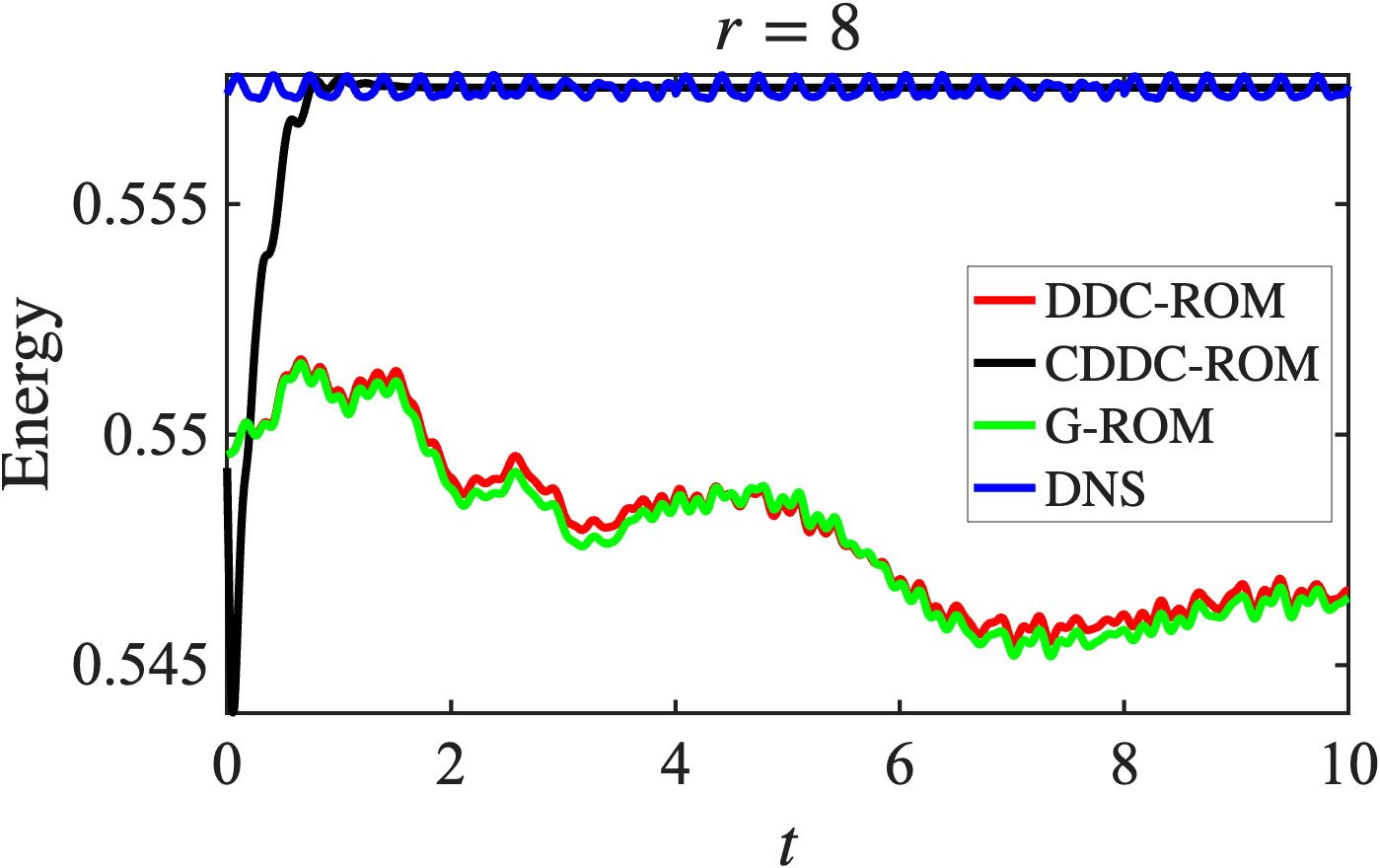}}\vspace{-4ex}
	\caption{Performance of the in-sample ROMs for correlation length $\ell=0.1$: Energy evolution comparing the DNS, G-ROM, DDC-UQ-ROM, and CDDC-UQ-ROM for (a) $r=5$ with $tol_d=8.0\times10^{-2}$, $tol_c=2.5\times10^{-2}$, and $\epsilon=3.6\times10^{-2}$; (b) $r=6$ with $tol_d=3.0\times10^{-4}$, $tol_c=3.0\times10^{-2}$, and $\epsilon=7.0\times10^{-2}$; (c) $r=7$ with $tol_d=4.0\times10^{-2}$, $tol_c=5.0\times10^{-1}$, and $\epsilon=1.0\times10^{-1}$; and (d) $r=8$ with $tol_d=5.0\times10^{-1}$, $tol_c=5.0\times10^{-1}$, and $\epsilon=1.0\times10^{-2}$.}\label{l=0.1-rom}\vspace{-4ex} 
\end{figure}

In this section, we investigate the performance of the G-ROM, DDC-UQ-ROM, and CDDC-UQ-ROM for an increased correlation length $\ell$. A larger value of $\ell$ results in a viscosity field with stronger spatial correlation and smoother spatial variations, while also affecting the magnitude of the viscosity fluctuations. To this end, we consider $\ell=0.1$ and a random Reynolds number with mean $Re=100$, and run the FOM to generate the DNS data for the in-sample ROM simulations. That is, in this case, DNS and ROMs all are computed using $N_{sc}=N_{train}=11$ five-dimensional collocation points and their corresponding weights. We compare the time evolution of the mean kinetic energy obtained with the G-ROM, DDC-UQ-ROM, and CDDC-UQ-ROM for several values of $r$ in Fig. \ref{l=0.1-rom}.

For $r=5$, the DDC-UQ-ROM and CDDC-UQ-ROM operators are constructed using $tol_d=8\times10^{-2}$ and $tol_c=2.5\times10^{-2}$, $\epsilon=3.6\times10^{-2}$, respectively. The performance of the ROMs is compared in Fig. \ref{l=0.1-rom} (a). We observe that the DDC-UQ-ROM does not provide a noticeable improvement over the G-ROM. In contrast, the CDDC-UQ-ROM yields significantly more accurate results and provides a better approximation of the DNS.

The performance of the ROMs for $r=6,7,$ and $8$ is presented in Fig. \ref{l=0.1-rom} (b), (c), and (d), respectively, along with the corresponding values of $tol_d$, $tol_c$, and $\epsilon$. In all three cases, the G-ROM and DDC-UQ-ROM exhibit relatively large fluctuations. Based on their time evolution, it is difficult to determine whether the DDC-UQ-ROM performs better than the G-ROM, as neither model reaches a statistically steady state. In contrast, the CDDC-UQ-ROM approximates the DNS results with significantly greater accuracy. In particular, for $r=8$, following an initial spin-up period, the CDDC-UQ-ROM exhibits excellent agreement with the DNS results. Overall, for the correlation length $\ell=0.1$, the CDDC-UQ-ROM consistently outperforms the other ROMs considered. For 801 sample five-dimensional collocation points, which is the out-of-sample case, the FOM blows up, thus we could not investigate the ROM performance.

\subsection{Efficiency of the Data-driven ROMs}

In this section, we compare the computational efficiency of the FOM, G-ROM, DDC-UQ-ROM, and CDDC-UQ-ROM in terms of computing time. To this end, we consider $\ell=0.01$, random Reynolds number with mean $Re=100$ and record the simulation time for the FOM with $N_{sc}=801$, as well as for the out-of-sample G-ROM, DDC-UQ-ROM, and CDDC-UQ-ROM.

\subsubsection{FOM computational cost}
The FOM computations were carried out in MATLAB R2023a on the Cheaha high-performance computing cluster at UAB. Simulations corresponding to the $N_{sc}=801$ stochastic collocation points 
were executed as a Slurm job array and distributed across 
\texttt{x86\_64} Intel- and AMD-based compute nodes. That is, depending on node availability, many jobs ran simultaneously. Each realization 
was allocated 2 CPU cores and 32~GB of memory. The total wall-clock time, 
measured from the start of the first array task to the completion of the 
last task, was 296,172 s.

\subsubsection{ROM computational cost}

All ROM computations were performed in MATLAB R2026a on a computer equipped with an Apple M3 Max chip and 64~GB of unified memory. 

For the ROMs, we report the offline and online computational times for $r=5$ and $d=16$. All offline and online codes are run in serial mode. The offline cost includes the construction of the ROM basis, ROM matrices, and the data-driven closure operators for the DDC-UQ-ROM and CDDC-UQ-ROM. In the offline stage, the ROM basis are computed for $N_{train}=11$ stochastic collocation points.

To model the subfilter-scale stress tensor in \eqref{sub-filter-scale-stress-tensor} using a full period of snapshot data, corresponding to $\mathbb{E}[\nu]=10^{-3}$ and consisting of a total of 166 snapshots, we determine the optimal data-driven ROM closure operators by solving \eqref{time-uncos} for the DDC-UQ-ROM and \eqref{const-optimi-data} for the CDDC-UQ-ROM. The computational times required to determine  optimal $\widetilde{\A}$ and $\widetilde{\B}$ for the DDC-UQ-ROM and $\widehat{\A}$ and $\widehat{\B}$ for the CDDC-UQ-ROM are 34.07 s  and 34.20 s, respectively. The total computational times required by the FOM and the ROMs, including both their offline and online stages, are reported in Table \ref{tab:computational_time_r5}.

\begin{table}[htbp]
\centering
\begin{tabular}{lcccc}
\hline
Model & Offline Time (s) & Online Time (s) 
& Total Time (s) & Wall-clock ratio \\
\hline
FOM      & --      & --     & 296172.00 & 1.00 \\
G-ROM    & 1130.38 & 49.14  & 1179.52   & 251.09 \\
DDC-UQ-ROM  & 1164.45 & 105.29 & 1269.74   & 233.25 \\
CDDC-UQ-ROM & 1164.58 & 104.26 & 1268.84   & 233.42 \\
\hline
\end{tabular}
\caption{Computational times for the FOM and out-of-sample ROMs with $r=5$ and $N_{sc}=801$ realizations. The wall-clock ratio is defined as the recorded FOM wall-clock time divided by the corresponding ROM total computational time. Since the FOM and ROM computations were performed on different computing platforms, these ratios are intended to provide an indicative measure of computational cost reduction rather than a hardware-normalized speedup.}
\label{tab:computational_time_r5}
\end{table}

 We emphasize that the FOM and ROM computations were performed on different computing platforms. Therefore, the computational times reported in Table~\ref{tab:computational_time_r5} should not be interpreted as a hardware-normalized performance benchmark. Rather, they provide an indicative comparison of the wall-clock computational costs associated with the FOM and ROM UQ approaches in the computing environments used in this study.

Despite this difference in hardware, the results clearly illustrate the substantial reduction in computational cost enabled by the ROMs. For the considered out-of-sample problem with $N_{sc}=801$ stochastic collocation points, the total recorded wall-clock time is $296172$ s for the FOM, compared with approximately $1270$ s for the DDC-UQ-ROM and CDDC-UQ-ROM, corresponding to an indicative wall-clock reduction factor of approximately $233$. The primary purpose of this comparison is therefore to demonstrate the practical computational savings obtained by replacing repeated high-fidelity simulations with the proposed reduced-order models, rather than to provide a hardware-independent speedup measurement.

\subsection{Sensitivity to the Amount of Closure-Training Data}

To investigate the effect of the amount of training data on the performance of the DDC-UQ-ROM and CDDC-UQ-ROM, we consider two reduced-data studies. In all experiments, we use the out-of-sample ROM setting with correlation length $\ell=0.01$ for the problem with a random Reynolds number having mean $Re=100$. Recall that the snapshots generated by the FOM \eqref{eq:FOM_momentum-expec}--\eqref{eq:FOM_continuity-expec} with $\mathbb{E}[\nu]=10^{-3}$ at times $t_{k^*+j}$, $k^*=3500$, $1\le j\le 166$, span one complete period. In contrast to the results obtained using all 166 snapshots, the studies below use only a limited subset of these snapshots to construct the data-driven closure operators. The purpose is to assess how strongly the performance of the proposed ROMs depends on the amount and temporal distribution of the closure-training data.

In the first study, we use only the first 11 snapshots, corresponding to the time levels $t_{k^*+j}$, $1\le j\le 11$, to construct the ROM closure models. Thus, these snapshots contain information only from the initial portion of the period and no information from the later portion. They are used in the unconstrained and constrained optimization problems to determine the ROM operators $\widetilde{\A}$, $\widetilde{\B}$, $\widehat{\A}$, and $\widehat{\B}$. These 11 snapshots represent approximately $6.6\%$ of the available data over one complete period. We refer to this as the limited, unequally spaced snapshot case.

\begin{figure}
	\centering
	\subfloat[]{\includegraphics[width=0.45\textwidth,height=0.3\textwidth]{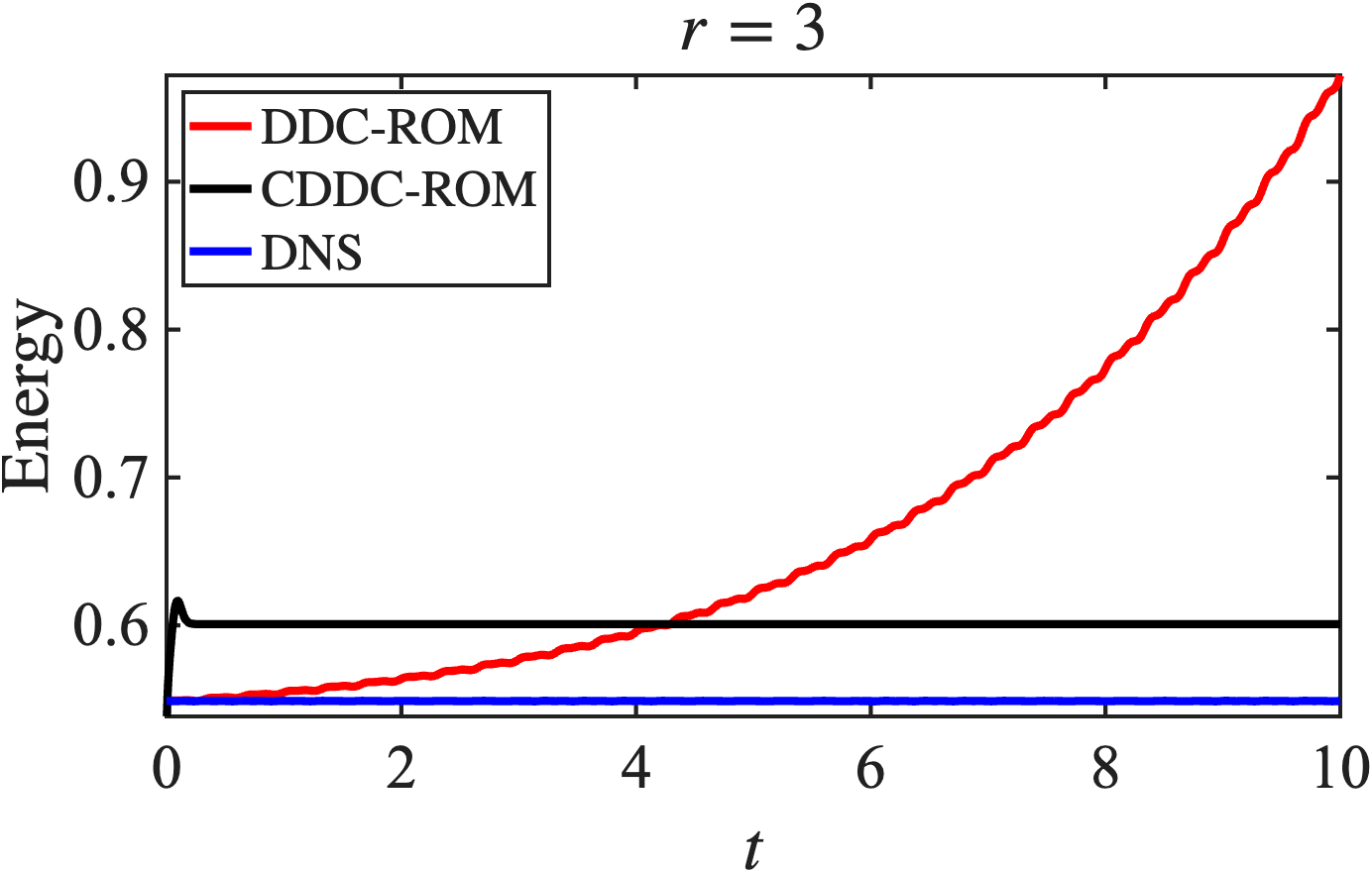}}\hspace{2ex}
    \subfloat[]{\includegraphics[width=0.45\textwidth,height=0.3\textwidth]
    {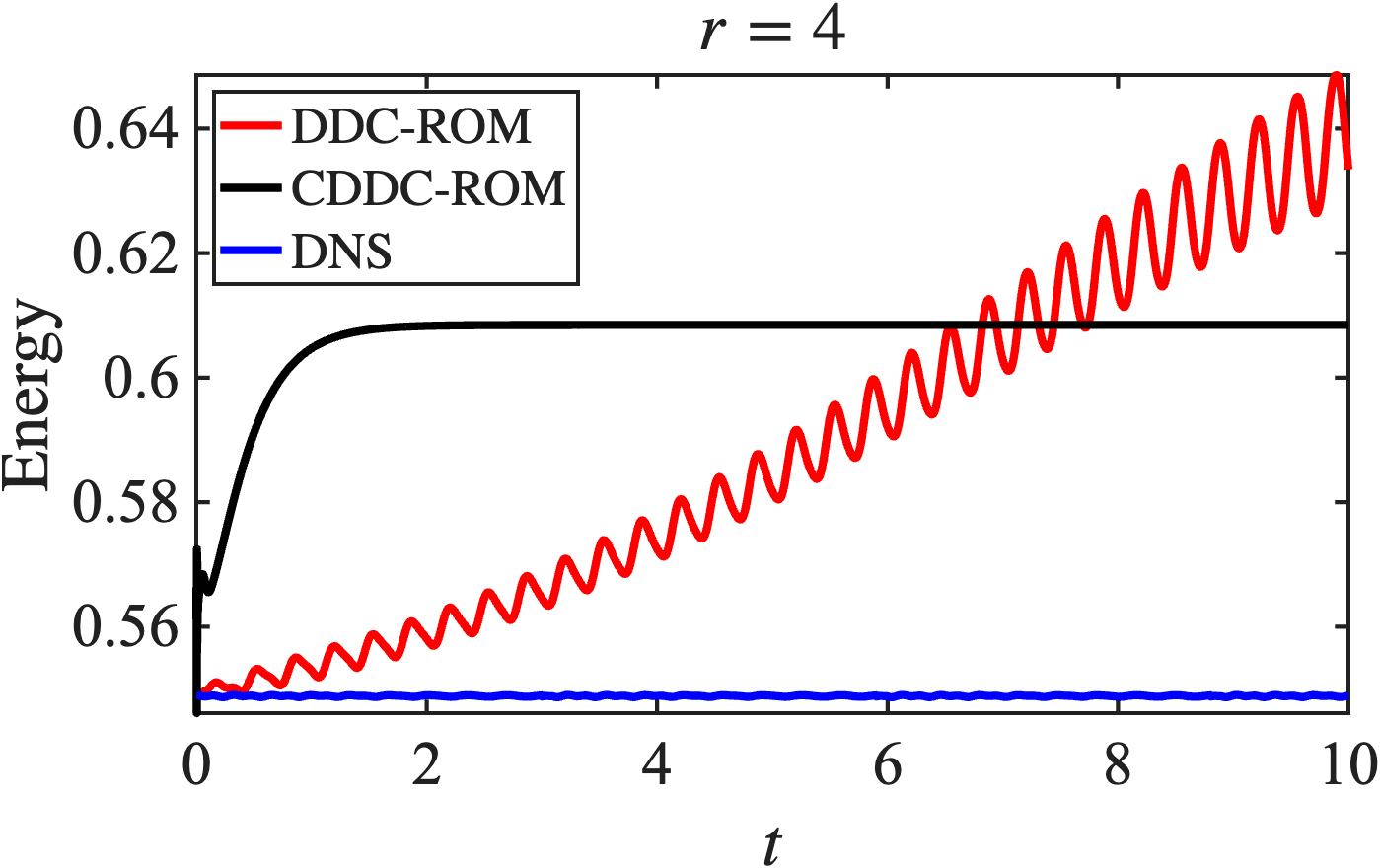}}\vspace{-4ex}
	\caption{Performance of the DDC-UQ-ROM and CDDC-UQ-ROM using only $6.6\%$ of the available closure-training snapshots from one complete period for flow past a circular cylinder with a random Reynolds number having mean $Re=100$. The selected snapshots are taken from the initial portion of the period. Energy evolution is shown for the DDC-UQ-ROM, CDDC-UQ-ROM, and DNS for (a) $r=3$ and (b) $r=4$.}\label{cross-val-11}
\end{figure}

In Fig.~\ref{cross-val-11}, we compare the DDC-UQ-ROM and CDDC-UQ-ROM results with the DNS for (a) $r=3$ with $tol_d=3\times10^{-6}$, $tol_c=4\times10^{-2}$, and $\epsilon=7\times10^{-4}$, and (b) $r=4$ with $tol_d=3\times10^{-2}$, $tol_c=4\times10^{-2}$, and $\epsilon=5.6\times10^{-3}$. Comparing the out-of-sample results in Fig.~\ref{r=3_T=10_wno_801} (b), obtained using the full closure-training dataset, with those in Fig.~\ref{cross-val-11} (a), obtained using only $6.6\%$ of the available snapshots, we observe a noticeable deterioration in the accuracy of both the DDC-UQ-ROM and CDDC-UQ-ROM. Nevertheless, the CDDC-UQ-ROM remains substantially more accurate than the DDC-UQ-ROM. These results indicate that the physically constrained closure retains a clear advantage even when only a small amount of training information is available. They also demonstrate that the amount of information contained in the projected subfilter-scale stress tensor has a significant influence on the accuracy and dissipative behavior of the resulting data-driven ROMs.

In the second study, we use 34 snapshots at the time levels $t_{k^*+5j-4}$, $1\le j\le 34$, generated by the FOM \eqref{eq:FOM_momentum-expec}--\eqref{eq:FOM_continuity-expec} with $\mathbb{E}[\nu]=10^{-3}$. These snapshots are used in the unconstrained and constrained optimization problems to construct the data-driven ROM closure operators $\widetilde{\A}$, $\widetilde{\B}$, $\widehat{\A}$, and $\widehat{\B}$. Thus, approximately $20.48\%$ of the available snapshot data over one complete period are used for closure training. Unlike the first study, the selected snapshots are uniformly distributed over the complete period; we therefore refer to this as the limited, equally spaced snapshot case.

\begin{figure}
	\centering
	\subfloat[]{\includegraphics[width=0.45\textwidth,height=0.3\textwidth]{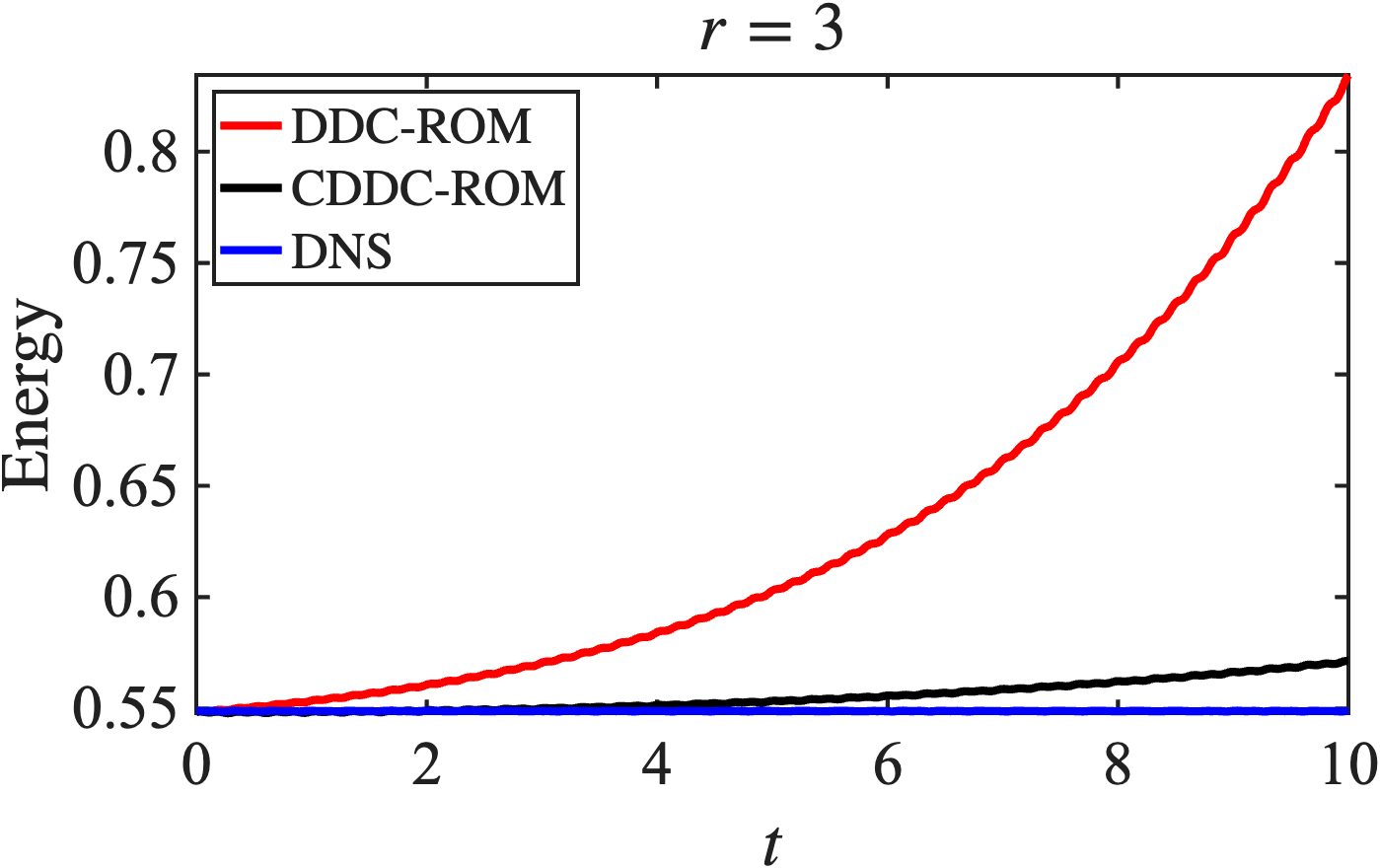}}\hspace{2ex}
    \subfloat[]{\includegraphics[width=0.45\textwidth,height=0.3\textwidth]
    {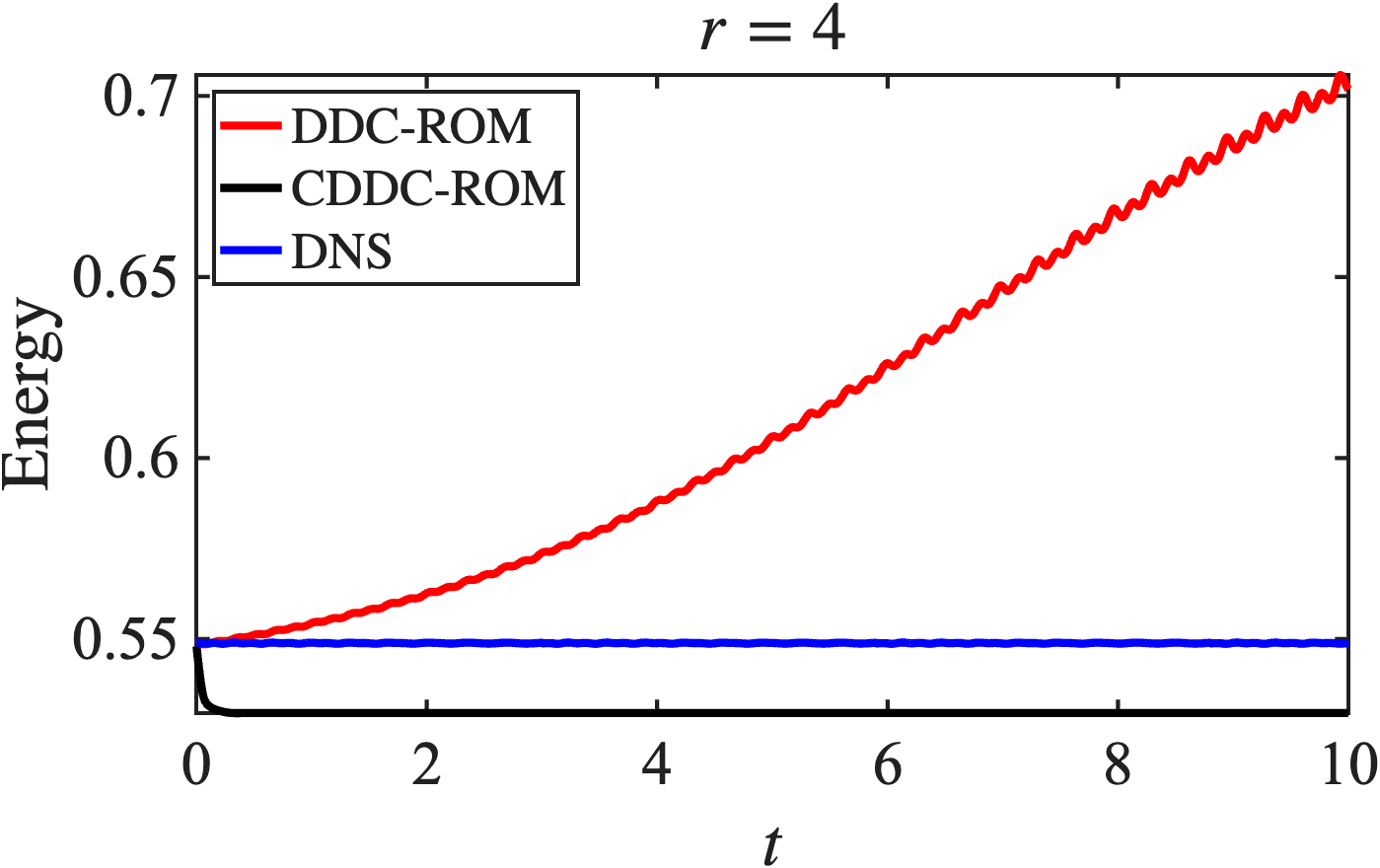}}\vspace{-4ex}
	\caption{Performance of the DDC-UQ-ROM and CDDC-UQ-ROM using $20.48\%$ of the available closure-training snapshots from one complete period for flow past a circular cylinder with a random Reynolds number having mean $Re=100$. The selected snapshots are uniformly distributed over the complete period. Energy evolution is shown for the DDC-UQ-ROM, CDDC-UQ-ROM, and DNS for (a) $r=3$ and (b) $r=4$.}\vspace{-4ex}\label{cross-val-alt-11}
\end{figure}

In Fig.~\ref{cross-val-alt-11}, we compare the energy evolution obtained with the DDC-UQ-ROM and CDDC-UQ-ROM against the DNS for (a) $r=3$ with $tol_d=9\times10^{-3}$, $tol_c=7\times10^{-1}$, and $\epsilon=6\times10^{-10}$, and (b) $r=4$ with $tol_d=4\times10^{-3}$, $tol_c=7\times10^{-1}$, and $\epsilon=3.5\times10^{-2}$. In both cases, the CDDC-UQ-ROM provides a more accurate approximation of the DNS energy than the DDC-UQ-ROM. Compared with the first reduced-data study, in which only $6.6\%$ of the available snapshots are used, the use of a larger and more broadly distributed subset of the closure-training data substantially improves the ability of the data-driven ROMs to reproduce the DNS energy evolution. For $r=4$, the CDDC-UQ-ROM exhibits slightly greater energy dissipation than the DNS, which indicates that the reduced amount of subfilter-scale stress information still influences the learned closure. Nevertheless, both data-driven ROMs provide good approximations of the DNS, with the physically constrained CDDC-UQ-ROM consistently yielding the better agreement.

Taken together, the two reduced-data studies demonstrate that the accuracy of the proposed data-driven closure models depends on both the amount and temporal distribution of the available training data. They also show that imposing the physical constraints in the CDDC-UQ-ROM substantially improves robustness when the closure operators are constructed from only a fraction of the full-period snapshot data.

\subsection{Statistical Error Analysis of the ROMs}

To investigate the sensitivity of the ROM models, in this section we compute errors in the time evolution of energy with respect to the first- and second-order moments for various values of the ROM dimension $r$. We consider the problem with a random Reynolds number having mean $Re=100$ and $\ell=0.01$, as described in Section \ref{cor-len-0-01}. We also use the same hyperparameters employed in generating Fig.~\ref{r=3_T=10_wno_801}--\ref{r=8_T=10_wno_801}. The time-averaged errors in the first- and second-order moments for the in-sample parameters are reported in Tables~\ref{in-sample-1st-order-mom} and \ref{in-sample-2nd-order-mom}, respectively, while those for the out-of-sample parameters are reported in Tables~\ref{out-sample-1st-order-mom} and \ref{out-sample-2nd-order-mom}, respectively.

For the first-order moment, we observe that the data-driven ROMs consistently outperform the G-ROM, with the CDDC-UQ-ROM providing the most accurate estimates. For the second-order moment, an exception occurs at $r=5$ for both the in-sample and out-of-sample cases, where the DDC-UQ-ROM outperforms the G-ROM, which in turn outperforms the CDDC-UQ-ROM. In all other cases, the CDDC-UQ-ROM provides the most accurate second-order moment estimates, followed by the DDC-UQ-ROM and then the G-ROM.

\begin{table}[h!]
\centering
\footnotesize
\begin{tabular}{|c|c|c|c|c|c|c|}
\hline
 $r$ & $3$ & $4$ & $5$  & $7$ & $8$ \\
\hline

 G-ROM
& $1.8102\times10^{-1}$ & $8.5116\times10^{-2}$ & $9.1589\times10^{-3}$
 & $2.0507\times10^{-2}$ & $2.1661\times10^{-2}$ \\
\hline

 DDC-UQ-ROM
& $9.2252\times10^{-2}$ & $5.6673\times10^{-2}$ &$7.4796\times10^{-3}$
 & $1.5969\times10^{-2}$ & $1.9581\times10^{-2}$ \\
\hline

CDDC-UQ-ROM
& $5.0896\times10^{-3}$ & $4.0200\times10^{-3}$ & $2.3270\times10^{-4}$
 & $1.2372\times10^{-3}$ & $2.1980\times10^{-3}$ \\
\hline
\end{tabular}

\caption{Time-averaged errors in the first-order energy moment for the in-sample parameters.}
\label{in-sample-1st-order-mom}
\end{table}

\begin{table}[h!]
\centering
\footnotesize
\begin{tabular}{|c|c|c|c|c|c|}
\hline
 $r$ & $3$ & $4$ & $5$  & $7$ & $8$ \\
\hline

 G-ROM
& $7.2039\times10^{-2}$& $6.7708\times10^{-3}$ & $4.4027\times10^{-6}$&   $1.7481\times10^{-3}$ &$1.9832\times10^{-3}$\\\hline

 DDC-UQ-ROM
&$2.1764\times10^{-2}$ & $4.7999\times10^{-3}$ &$1.7106\times10^{-6}$ &  $1.1831\times10^{-3}$ &$1.6735\times10^{-3}$\\\hline

CDDC-UQ-ROM
& $1.3186\times10^{-4}$& $1.3854\times10^{-4}$ & $5.0693\times10^{-5}$& $1.5164\times10^{-4}$  &$1.9983\times10^{-5}$\\\hline
\end{tabular}

\caption{Time-averaged errors in the second-order energy moment for the in-sample parameters.}
\label{in-sample-2nd-order-mom}
\end{table}

\begin{table}[h!]
\centering
\footnotesize
\begin{tabular}{|c|c|c|c|c|c|}
\hline
 $r$ & $3$ & $4$ & $5$  & $7$ & $8$ \\
\hline

 G-ROM
 & $1.6687\times10^{-1}$ & $8.5637\times10^{-2}$ & $8.9734\times10^{-3}$&  $1.7905\times10^{-2}$& $2.1629\times10^{-2}$ \\\hline

 DDC-UQ-ROM
& $8.3374\times10^{-2}$ & $5.5487\times10^{-2}$ & $7.2680\times10^{-3}$& $1.4008\times10^{-2}$ & $1.9630\times10^{-2}$ \\\hline

CDDC-UQ-ROM
& $5.9622\times10^{-3}$ & $4.2880\times10^{-3}$ & $4.6749\times10^{-4}$& $1.8087\times10^{-3}$ & $1.9630\times10^{-3}$ \\\hline
\end{tabular}

\caption{Time-averaged errors in the first-order energy moment for the out-of-sample parameters.}
\label{out-sample-1st-order-mom}
\end{table}

\begin{table}[h!]
\centering
\footnotesize
\begin{tabular}{|c|c|c|c|c|c|c|}
\hline
 $r$ & $3$ & $4$ & $5$  & $7$ & $8$ \\
\hline

 G-ROM
& $5.9381\times10^{-2}$&  $5.2211\times10^{-3}$& $6.7948\times10^{-6}$&  $1.9791\times10^{-3}$ &$2.2070\times10^{-3}$\\\hline

 DDC-UQ-ROM
& $1.8221\times10^{-2}$& $3.7179\times10^{-3}$ &$4.5945\times10^{-6}$ &  $1.3907\times10^{-3}$ &$1.8728\times10^{-3}$\\\hline

CDDC-UQ-ROM
& $1.0914\times10^{-4}$& $6.6142\times10^{-5}$ & $3.6903\times10^{-5}$&  $1.7127\times10^{-4}$ &$2.2398\times10^{-5}$\\\hline
\end{tabular}

\caption{Time-averaged errors in the second-order energy moment for the out-of-sample parameters.}
\label{out-sample-2nd-order-mom}
\end{table}

\begin{remark}
    The hyperparameters \(tol_d\), \(tol_c\), and \(\epsilon\) used in the DDC-UQ-ROM and CDDC-UQ-ROM were selected empirically through numerical experimentation to illustrate the potential accuracy of the proposed data-driven reduced-order models. In the present study, the DNS results were used as reference solutions in selecting parameter values that provide accurate ROM approximations. Thus, the objective of these experiments is to assess the achievable performance of the proposed closure models rather than to develop an automated hyperparameter-selection procedure. In practical applications, these parameters could instead be determined using a separate training/validation dataset or an automated parameter-search procedure. The development and assessment of such hyperparameter-selection strategies are beyond the scope of the present work.
\end{remark}

\section{Conclusion and Future Directions}\label{conclusion}

In this work, we developed two hybrid data-driven filtered correction ROMs for UQ of the time-dependent incompressible NSEs with random viscosity. We introduced a stochastic formulation in which the realization-dependent closure term $\btau_j$ is approximated by the closure associated with the expected viscosity, $\btau_j\approx\btau_{\bar{\nu}}$. This approximation allows a single data-driven closure model, constructed from the FOM corresponding to $\mathbb{E}[\nu]$, to be reused across all realizations and thereby avoids the prohibitive cost of constructing a separate closure model for each realization. The resulting DDC-UQ-ROM and its physically constrained variant (CDDC-UQ-ROM) were combined with the stochastic collocation method to provide an efficient nonintrusive framework for UQ.

The proposed methods were investigated for 2D flow past a circular cylinder with a 5D random viscosity field. Numerical experiments were performed for random $Re$ with mean values $Re=100$ and $Re=500$, and for different $\ell$ of the random viscosity. Both in-sample and out-of-sample parameter settings were considered. In particular, the ROM basis was constructed using $N_{\mathrm{train}}=11$ stochastic collocation points, while the out-of-sample calculations employed $N_{\mathrm{sc}}=801$ collocation points. For the cases with correlation length $\ell=0.01$ and mean $Re=100$, the DDC-UQ-ROM generally improves upon the G-ROM, whereas the CDDC-UQ-ROM provides the most accurate approximation of the DNS energy evolution over the range of ROM dimensions investigated. For the more convection-dominated case with mean $Re=500$, the CDDC-UQ-ROM again provides the closest agreement with the DNS in both the in-sample and out-of-sample experiments. For the larger correlation length $\ell=0.1$, the CDDC-UQ-ROM also yields substantially better agreement with the available in-sample DNS results than the G-ROM and DDC-UQ-ROM.

These results also provide a posteriori support for the approximation $\btau_j\approx\btau_{\bar{\nu}}$. Although the data-driven correction operators are constructed from FOM information corresponding only to $\mathbb{E}[\nu]$, the resulting ROMs are subsequently applied over a much larger set of stochastic parameter realizations. In particular, the favorable out-of-sample results obtained using 801 stochastic collocation points indicate that the expected-viscosity closure is able to capture the dominant unresolved-scale effects over the parameter range considered without requiring realization-specific closure models.

The statistical error analysis further demonstrates the effectiveness of the proposed approach. For the first-order moment of the kinetic energy, the data-driven ROMs consistently improve upon the G-ROM, with the CDDC-UQ-ROM yielding the smallest time-averaged errors for all ROM dimensions considered. For the second-order moment, the CDDC-UQ-ROM is also the most accurate method in nearly all cases; the exception occurs at $r=5$, where the DDC-UQ-ROM gives the smallest error, followed by the G-ROM and then the CDDC-UQ-ROM. Overall, these results show that the physical constraints substantially improve the ability of the data-driven closure model to reproduce not only the mean energy evolution but also its statistical behavior.

The reduced-training-data experiments provide additional information about the robustness of the closure models. When only $6.6\%$ of the available closure-training snapshots from the initial portion of one complete period are used, the accuracy of both data-driven ROMs deteriorates, but the CDDC-UQ-ROM remains substantially more accurate than the DDC-UQ-ROM. Increasing the training information to $20.48\%$ of the available snapshots and distributing those snapshots over the complete period significantly improves the ROM approximations. These results demonstrate that both the amount and temporal distribution of the closure-training data influence the learned reduced-order dynamics and that the physical constraints incorporated into the CDDC-UQ-ROM improve robustness when the available training information is limited.

The computational results demonstrate the potential efficiency of the proposed framework for many-query uncertainty quantification. For the out-of-sample problem with $N_{\mathrm{sc}}=801$ stochastic collocation points, the recorded FOM wall-clock time was $296172$ s, whereas the total offline and online costs of the DDC-UQ-ROM and CDDC-UQ-ROM were approximately $1270$ s. Because the FOM and ROM computations were performed on different computing platforms, this comparison should not be interpreted as a hardware-normalized benchmark; nevertheless, it corresponds to an indicative wall-clock reduction factor of approximately $233$ and illustrates the substantial computational savings that can be obtained by replacing repeated high-fidelity simulations with the proposed reduced-order models.

Several directions remain for future investigation. First, the hyperparameters $tol_d$, $tol_c$, and $\epsilon$ were selected empirically in the present study in order to investigate the achievable performance of the proposed closure models. Developing systematic and automated strategies for selecting these parameters using independent training and validation data would further improve the predictive use of the methods. Second, a systematic comparison with other stabilized ROM UQ approaches, such as the evolve-filter-relax stochastic collocation ROM \cite{gunzburger2019evolve}, would help quantify the relative advantages of different stabilization and closure strategies. Additional investigations involving higher Reynolds numbers, more complex stochastic parameterizations, and other quantities of interest would further assess the robustness and generality of the proposed framework. Finally, the data-driven correction methodology is not specific to the Navier--Stokes equations and can be extended to other nonlinear stochastic PDEs for which repeated high-fidelity simulations make UQ computationally prohibitive.

\section{Acknowledgment} The authors are grateful to the NSF
for supporting this work under Grant DMS-2425308 and to the
Alabama Supercomputer Authority (ASA) for providing access
to high-performance computing resources. The authors thank Traian Iliescu and Leo G. Rebholz for sharing their expertise and valuable insights.

\bibliography{rom2}
\end{document}